\documentclass[aps,prb,onecolumn,a4paper,11pt,showpacs,citeautoscript,notitlepage,preprintnumbers]{revtex4-1} 

\usepackage{graphicx,amsfonts,epsfig,latexsym,amscd,amsmath,theorem,mathrsfs,color}
\usepackage[colorlinks=true, a4paper=true, pdfstartview=FitV,linkcolor=blue, citecolor=blue, urlcolor=blue]{hyperref}
\usepackage{bm}

\usepackage[markup=default]{changes}

\def\comment#1{}

\newcommand{\beg}{\begin{eqnarray}}
\newcommand{\eee}{\end{eqnarray}}

\def\cm#1{}

\newcommand{\R}{{\mathbb{R}}}

\newcommand{\hh}{{\cal H}}
\newcommand{\eps}{\epsilon}

\newcommand{\f}{\frac}

\newcommand{\be}{\begin{equation}}
\newcommand{\ee}{\end{equation}}
\newcommand{\ba}{\begin{eqnarray}}
\newcommand{\ea}{\end{eqnarray}}
\newcommand{\beq}{\begin{equation}}
\newcommand{\eeq}{\end{equation}}
\newcommand{\bea}{\begin{eqnarray}}
\newcommand{\eea}{\end{eqnarray}}
\newcommand{\bastar}{\begin{eqnarray*}}
\newcommand{\eastar}{\end{eqnarray*}}

\newcommand{\cd}{\partial}

\newcommand{\ignore}[1]{}
\newcommand{\mbf}{\mathbf}
\newcommand{\Eqref}[1]{\eqref{#1}}
\newcommand{\Figref}[1]{Fig.~\ref{#1}}

\newcommand{\ie}{{\it i.e.~}}
\newcommand{\resp}{\emph{resp.~}}

\begin{document}
\preprint{Physica C : Superconductivity {\bf  533}, 20-35 (2017)}
\title{\texorpdfstring{Type-1.5 superconductivity in multicomponent systems. }{Type-1.5 superconductivity in multicomponent systems.}}

\author{
E. Babaev${}^{1}$, J. Carlstr\"om${}^{2}$,  M. Silaev${}^{1}$ and  J.M. Speight${}^3$}
\affiliation{
${}^1$Department of Theoretical Physics and Center for Quantum Materials, The Royal Institute of Technology, Stockholm, SE-10691 Sweden\\
${}^2$ Department of Physics, University of Massachusetts Amherst, MA 01003 USA\\
${}^3$ School of Mathematics, University of Leeds, Leeds LS2 9JT, UK
}

\begin{abstract}
In general a superconducting state breaks multiple symmetries and, therefore, is
 characterized by several different coherence lengths $\xi_i$, $i=1,...,N$.
 Moreover in multiband material even 
 superconducting states  that break
only a single symmetry  are nonetheless described, under certain conditions by multi-component theories
with multiple coherence lengths.
As a result of that there can appear
 a   state 
  where some coherence lengths
  are larger and some are smaller than the magnetic field penetration length $\lambda$: $\xi_1\leq \xi_2... < \sqrt{2}\lambda<\xi_M\leq...\xi_N$. That state was recently
  termed ``type-1.5" superconductivity. 
 This breakdown of type-1/type-2 dichotomy  is  rather generic near a phase transition 
  between superconducting states with different symmetries.
  The examples include the  transitions between
  $U(1)$ and $U(1)\times U(1)$ states 
  or  between $U(1)$ and $U(1)\times Z_2$ states. The later example is realized
  in systems that feature transition  between $s$-wave and $s+is$ states.
The extra fundamental length scales have many physical consequences.  In particular in these regimes vortices can attract one another at long range but repel at shorter ranges.
Such a system can form vortex clusters in low magnetic fields.
The vortex clustering in the type-1.5 regime gives rise to many physical effects, ranging from 
macroscopic phase separation in domains of different broken symmetries, to unusual
 transport properties.
 {\sl Prepared for the proceedings of Vortex IX  conference, Rhodes 12-17 September 2015}.
\end{abstract}
\maketitle
\section{Introduction}
Below we briefly discuss the properties and occurrence of ``type-1.5" superconducting state characterized by multiple coherence length, some of which are smaller and some larger than the magnetic field penetration length \cite{Babaev.Speight:05}  $\xi_1\leq \xi_2... < \sqrt{2}\lambda<\xi_M\leq...\xi_N$.
Type-1 superconductors expel weak magnetic fields, while strong fields give rise to formation of macroscopic phase separation in the form of domains of Meissner and normal states 
\cite{landau1950k,de1999superconductivity}. The response of type-2 superconductors is
the following \cite{abrikosov1957magnetic}: below some critical value $H_{c1}$, the field is expelled. Above this value a superconductor forms a lattice
or a liquid of vortices which carry magnetic flux through the system. Only at a higher second critical value, $H_{c2}$ is superconductivity destroyed.
These different responses are the consequences of  the form of the vortex interaction in these systems,
 the energy cost of a boundary between superconducting and normal states and  the thermodynamic stability
of vortex excitations. In a type-2 superconductor the
energy cost of a boundary between the normal and the superconducting state is
negative, while  the interaction between vortices is repulsive \cite{abrikosov1957magnetic}.
This leads to a formation of stable vortex lattices and liquids.
In type-1 superconductors the situation is the opposite; the vortex interaction is attractive (thus making them
unstable against collapse into one large ``giant" vortex),
while the boundary energy between normal and superconducting
states is positive.
The 'ordinary' Ginzburg-Landau model has a critical regime where 
 vortices do not interact \cite{kramer,Bogomol}. 
The critical value of $\kappa$
in the most common GL model parameterization corresponds to
 $\kappa = 1/\sqrt{2}$ (often the factor $1/\sqrt{2}$ is absorbed into the definition of coherence length 
 in which case the critical coupling is  $\kappa_c = 1$).  
The noninteracting regime, which is frequently called the ``Bogomolnyi limit''
is a property of Ginzburg-Landau model where, at $\kappa_c = 1/\sqrt{2}$,
the core-core attractive interaction between vortices   cancels at all distances
the current-current repulsive interaction \cite{kramer,Bogomol} (up to microscopic corrections
beyond the  standard Ginzburg-Landau theory).
By contrast in multi-component theories are characterized by multiple
coherence length and do not in general allow  to form  a single Ginzburg-Landau
parameter $\kappa$.

 The Ginzburg-Landau free energy functional for a multicomponent {superconductor} has the form

\begin{equation}
F=\frac{1}{2}\sum_i (D\psi_i)(D\psi_i)^* + V(|\psi_i|)
+\frac{1}{2}(\nabla\times {\bf A})^2 ,
\label{gl0}
\end{equation}
{where} $\psi_i$ are complex superconducting components,
$D=\nabla + ie {\bf A}$, and $\psi_i=|\psi_i|e^{i\theta_i}$,
$a=1,2$,  and $V(|\psi_i|)$ stands for effective potential.
{We consider a general form of} potential terms 
but the simplest gradient terms.{ In general  Eq.(\ref{gl0}) indeed can contain} mixed (with respect to components $\psi_i$)  gradient terms, e.g.
$Re [D_{\alpha=x,y,z}\psi_iD_{\beta=x,y,z}\psi_j]$
(for a more detail on the effects of these terms see \cite{johan2}).

The multiple  {superconducting} components can have various origins.
 { First of all they can} arise in {\it (i) superconducting states which break multiple symmetries}.
 { Such systems} are described by {several} order parameters in the sense of Landau theory 
 of phase transitions, and have  {different} coherence lengths associated with them. 
 Multiple broken symmetries are present even  in the simplest generalisation
of the s-wave superconducting states: the $s+is $ superconducting state 
 \cite{StanevTesanovic,Chubukov2},   {which} breaks $U(1)\times Z_2$ symmetry \cite{Johan}.
Likewise multiple broken symmetries are present in  non-s wave superconductors.
Another example is mixtures of independently conserved condensates such as the models for the theoretically discussed
superconductivity in metallic hydrogen and hydrogen rich alloys \cite{Nature,herland2010phase}.
There $\psi_i$ represents electronic and protonic Cooper pairs
or deuteronic condensates. A similar situation  was discussed in
certain models of nuclear superconductors in the interior of neutron stars, where $\psi_i$ represent protonic and $\Sigma^-$ hyperonic condensates  \cite{Jones21092006,babaev2009unconventional}.

{ Another class of multi-component superconductors is}  {\it (ii) systems which are 
described by  multi-component Ginzburg-Landau field {theories that do} not originate in multiple broken symmetries.}
The most common examples are multiband superconductors \cite{suhl,LeggettMode,tilley}. 
In this case  $\psi_i$ represent
superconducting components belonging to different bands.
{ Since a priory there are no symmetry constraints preventing interband Cooper pair tunnelling 
the theory contains generic terms which describe intercomponent Josephson coupling $\frac{\eta}{2}(\psi_i\psi_j^* +\psi_i\psi_j^*)$. These terms explicitly break symmetry.
 Here the number of components $\psi_i$ is not dictated by the broken symmetry pattern. 
}
 Multicomponent GL expansions can be justified when for example  $SU(N)$ or $[U(1)]^N$ symmetry is softly 
 explicitly {broken down} to $U(1)$ \cite{Silaev.Babaev:12}.  
  Some generalizations of type-1.5 concepts for the case of $p$-wave pairing in multiband systems was discussed in \cite{Garaud.Agterberg.ea:12}.
Recently rigorous mathematical work was done on justification of multicomponent Ginzburg-Landau expansions \cite{Frank2016}.


 \subsection{Type-1.5 superconductivity} 
Multicomponent systems allow a type of
superconductivity that is distinct from the type-1 and type-2 
 \cite{Babaev.Speight:05,johan1,johan2,Silaev.Babaev:11,nonpairwise,Johan,Silaev.Babaev:12,moshchalkov,moshchalkov2}. It
emerges from the following circumstances: 
{Multi-component GL models have several fundamental scales}, namely the
magnetic field penetration depth  $\lambda$ and multiple coherence lengths (characteristic
 scales of the variations of the density fields) $\xi_i$,
which renders the model impossible to parametrize in terms of a single
dimensionless parameter $\kappa$,
thus making the type-1/type-2 dichotomy insufficient for classifying and describing these systems.
Rather, in a wide range of parameters, there is a separate superconducting regime
with some coherence lengths that are larger and some that are smaller than the magnetic 
field penetration length $\xi_1/\sqrt{2}<\xi_2/\sqrt{2}<...<\lambda<\xi_M/\sqrt{2}<...<\xi_N/\sqrt{2}$. 
In that regime a situation is possible
where vortices have long-range attractive (due to ``outer cores" overlap) and short-range
repulsive interaction (driven by current-current and electromagnetic interaction)
and form vortex clusters coexisting with  domains of  two-component Meissner state  \cite{Babaev.Speight:05}.
The first experimental works  \cite{moshchalkov,moshchalkov2}
put forward that this
state is realized in
 the two-band material MgB$_2$. Moshchalkov et al termed  this regime
 ``type-1.5 superconductivity" \cite{moshchalkov}. 
Recently experimental works proposed that this state is realised in 
${\rm Sr_2Ru O_4}$ \cite{moler2,Ray.Gibbs.ea:14} and $\rm LaPt_3 Si$ \cite{noncentr,1347-4065-54-4-048001}.
The experiments on the MgB$_2$
were done using Bitter decoration, scanning SQUID and scanning Hall probes  \cite{moshchalkov,moshchalkov2}. The analysis of the nature of vortex clustering
was done by cycling field and observing vortex cluster formation in different parts of the sample.
In the case of ${\rm Sr_2Ru O_4}$  the evidence of intrinsic mechanism
of vortex cluster formation comes from $\mu SR$ experiment that observed
vortex clusters contraction with decreasing temperature well below $T_c$ thus
signaling attractive inter vortex forces rather than pinning
responsible for clusterization. The attractive inter vortex
forces also provides and explanation for earlier experiment
on ${\rm Sr_2Ru O_4}$  \cite{PhysRevB.65.144519}  that reported vanishing creep
in the absence of dramatic increase of critical current  [initially in \cite{PhysRevB.65.144519} this was  attributed to
effects of domain walls trapping vortices. However such configurations would have
very characteristic magnetic signatures \cite{garaud2016lattices,garaud2015properties}, these   signatures were not observed in surface probes \cite{moler2}].
A prediction of a (narrow)  region of type-1.5 state was made
for certain interface superconductors \cite{Agterberg.Babaev.ea:14}.
Also it was pointed out that a generic type-1.5 regime 
should form  in 
iron-based superconductors near transitions from $s$ to $s+is$ pairing states \cite{Johan}. 
Type-1.5 superconductivity was  discussed in
the context of the quantum Hall effect \cite{parameswaran2012typology} and neutron stars \cite{alford2008flux}.
For other recent works on this and related subjects see e.g.
\cite{dao,gutierrez2012scanning,li2011low,varney2013hierarchical,PhysRevB.82.132505,Reichhardt,meng2014honeycomb,Garaud.Agterberg.ea:12,Garaud.Babaev:15,edstrom2013three,forgacs2016vortices}.

In these systems one cannot straightforwardly use the usual one-dimensional
argument concerning the energy of superconductor-to-normal state boundary
to classify the magnetic response.
First of all, the energy per vortex in such a case
depends on whether a vortex is placed in a cluster or not.
Formation of a single isolated vortex might be energetically
unfavorable, while formation of vortex clusters
can be favorable, because in a cluster (where vortices are placed
in a minimum of the interaction potential), the energy
per flux quantum is smaller than that for an isolated vortex.
Besides the energy
of a vortex in a cluster, there appears
an additional characteristic associated
with the energy of the boundary of a cluster. In other words for systems
with inhomogeneous vortex states there are many different interfaces,
  some of which have positive and some  negative free energy. The
  non-monotonic intervortex interaction is one of key properties 
  of a type-1.5 superconductor, but is not a state-defining one.  
  As discussed in the introduction and also below,  intervortex attraction
  can arise under certain circumstances in single-component materials as well.
  In type-1.5 case  this is  a consequence of multiple coherence lengths and comes
  with a number of new physical effects discussed below.
We summarise the basic properties of type-1, type-2 and type-1.5 regimes in
the table \ref{table1} .

\begin{table*}
\begin{center}
\begin{tabular}{|p{3cm}||p{4.5cm}|p{4.5cm}|p{5cm}|}
\hline
 & {\bf  single-component type-1 } & {\bf single-component type-2}  & {\bf multi-component Type-1.5}  \\ \hline \hline
{\bf Characteristic lengths scales} & Penetration length $\lambda$ \&   coherence length $\xi$ ($\frac{\lambda}{\xi}< \frac{1}{\sqrt{2}}$) & Penetration length $\lambda$ \& coherence length $\xi$ ($\frac{\lambda}{\xi}> \frac{1}{\sqrt{2}}$) &Multiple characteristic density variations length scales $\xi_i$,  and penetration length $\lambda$,  the non-monotonic vortex interaction occurs in these systems in a large range of parameters when  $\xi_1\leq\xi_2\leq....<\sqrt{2}\lambda<\xi_M\leq ... \leq\xi_N$
\\ \hline
 {\bf  Intervortex interaction} &   Attractive &    Repulsive   &Attractive at long range and repulsive at short range \\ \hline
{\bf  Energy of superconducting/normal state boundary} &    Positive    & Negative   & Under quite general conditions negative energy of superconductor/normal interface inside a vortex cluster but positive energy  of the vortex cluster's boundary  \\ \hline
{\bf The magnetic field required to form a vortex} &    Larger than the thermodynamical critical magnetic field  & Smaller than thermodynamical critical magnetic field & In different cases either (i) smaller than the thermodynamical critical magnetic field or (ii) larger than critical magnetic field for single vortex but smaller than critical magnetic field for a vortex cluster of a certain critical size
\\ \hline
{\bf  Phases in external magnetic field } & (i) Meissner state at low fields; (ii) Macroscopically large normal domains at elevated fields.
First order phase transition between superconducting (Meissner) and normal states & (i) Meissner state at low fields, (ii) vortex lattices/liquids at larger fields.  Second order phase transitions between Meissner and vortex states and between vortex and normal states at the level of mean-field theory. & (i) Meissner state at low fields (ii) Macroscopic phase separation into vortex clusters coexisting with Meissner domains at intermediate fields (iii) Vortex lattices/liquids at larger fields.  Vortices form via a first order phase transition. The transition from vortex states to normal state is second order.
\\ \hline
{\bf  Energy E(N) of N-quantum axially symmetric vortex solutions} &    $\f{E(N)}{N}$  $<$ $\f{E(N-1)}{N-1}$ for all N. Vortices
collapse onto a single N-quantum mega-vortex &   $\f{E(N)}{N} >\f{E(N-1)}{N-1}$ for all N. N-quantum vortex
decays into N infinitely separated single-quantum vortices & There is a characteristic number N${}_c$ such
that $\f{E(N)}{N}$  $<$ $\f{E(N-1)}{N-1}$ for N $<$ N${}_c$, while $\f{E(N)}{N}$ $>$ $\f{E(N-1)}{N-1}$ for N
$>$ N${}_c$. N-quantum vortices decay into vortex clusters.
\\ \hline
\end{tabular}
\caption[Basic characteristics of superconductors]{Basic characteristics of bulk clean superconductors in type-1, type-2 and type-1.5 regimes. Here the most common units are used in which the value of the GL parameter
which separates type-1 and type-2 regimes  in a single-component theory is $\kappa_c=1/\sqrt{2}$.
Magnetization curves in these regimes are shown on Fig. \ref{magnetization}}
\label{table1}
\end{center}
\end{table*}


\begin{figure}
\includegraphics[width=\linewidth]{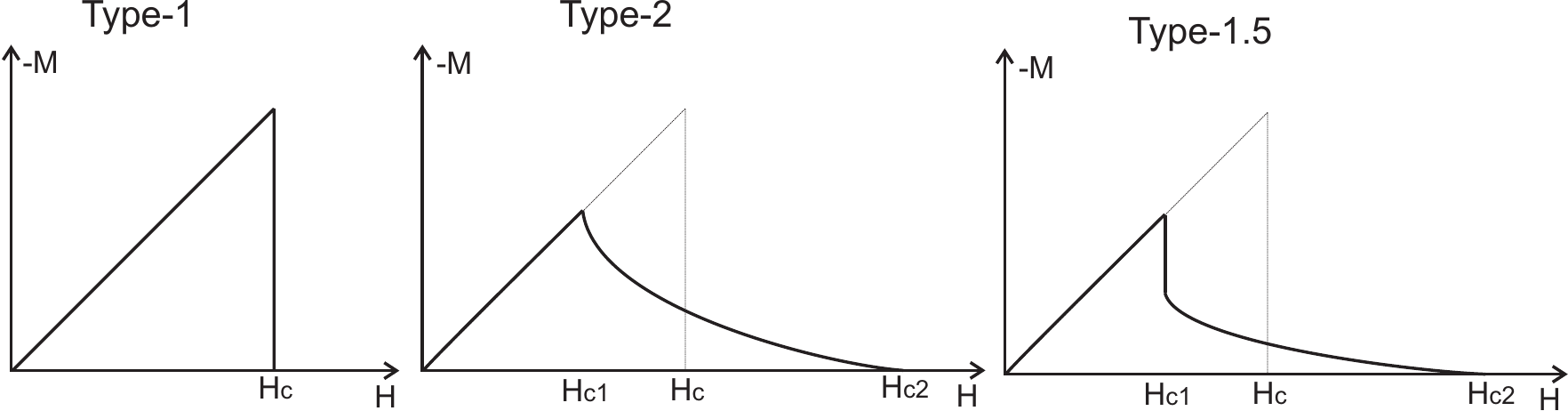}
\caption{A schematic picture of
magnetization curves of type-1, type-2 and type-1.5 superconductors. The magnetisation jump at $H_{c1}$ is one of the features of type-1.5 regime, however it is not a state-defining property conversion of the $H_{c1}$ phase transition
to a first order one can  be caused by a number of reasons (e.g. microscopic corrections near Bogomolnyi point, multi-layer structure, etc) 
in ordinary type-2 superconductors
}
\label{magnetization}
\end{figure}

\section{The two-band Ginzburg-Landau Model with   arbitrary interband
interactions. Definition of the coherence lengths
and type-1.5 regime}
\subsection{Free energy functional}

 { Realization of the type-1.5 regime requires at least two superconducting components.} 
In this section we study the type-1.5 regime using the following two-component Ginzburg-Landau (TCGL) free energy functional.

\begin{equation}
F=\frac{1}{2}(D\psi_1)(D\psi_1)^*+\frac{1}{2}(D\psi_2)(D\psi_2)^*
-\nu Re\Big\{(D\psi_1)(D\psi_2)^*\Big\}+\frac{1}{2}(\nabla\times {\bf A})^2 + F_p
\label{gl}
\end{equation}
Here
$D=\nabla + ie {\bf A}$, and $\psi_i=|\psi_i|e^{i\theta_i}$,
$i=1,2$, represent two superconducting  components.
While in general two components can have different critical temperatures, 
in the simplest case, the two-band superconductor breaks only
$U(1)$ symmetry. Then eq. (\ref{gl}) can be obtained 
as an expansion of the free energy in small gaps and small gradients
 \cite{tilley,gurevich,gurevich2,zhitomirsky,Silaev.Babaev:12,garaud2016microscopically}. Such an expansion
 should not be confused with the simplest expansion 
 in a single small parameter $\tau= (1-T/T_c)$. The   $\tau$-based expansion for $U(1)$ system
 is an approximation that
 yields only one order parameter for a $U(1)$ system and neglects 
 the second coherence length. The multi-parameter expansions that 
 are not based on symmetry  are
justified under certain conditions 
\cite{Silaev.Babaev:12,garaud2016microscopically}. Indeed the
existence of two bands in a superconductor by itself is {  not} a sufficient conditions for
a superconductor to be described by a model like (\ref{gl})
with two well-defined coherence lengths. 
For discussion of the applicability conditions of the 
theory (\ref{gl})  for two-band $U(1)$ systems see \cite{Silaev.Babaev:11,Silaev.Babaev:12}. 
Note that, in general in two-band expansion, the terms corresponding to one component can 
be larger than a  terms contributed by another component. However as it will be clear below, for the 
discussion of typology of superconductors, the relevant parameters are characteristic length scales
associated with the exponential laws at which field component restore their ground state values away from a perturbation such as a vortex core (i.e. the coherence lengths).
Indeed a component with {\it smaller amplitude} can give raise to a {\it longer  coherence length}  that is important for intervortex interaction and
should not be discarded based merely on the smallness of amplitude $|\psi_i |$.
In principle, for the   component with larger amplitude, one can keep higher-power terms in the GL expansion
such as $\psi_i\psi_i^*(D\psi_i)(D\psi_i)^*, |\psi_i|^6$ etc. These terms lead to some corrections to two coherence lengths,
while not affecting the overall form of intervortex
  forces as can be seen from the analysis    in \cite{johan2}. As be  seen from the  comparison of vortex solutions in the GL formalism and in microscopic model
  without GL expansion \cite{Silaev.Babaev:12}, in the regime of most interest these terms can be neglected.
  
 We begin  with {the} most general analysis by considering
 the case where $F_p$ can contain an {\it arbitrary} collection of non-gradient terms,
of arbitrary power in $\psi_i$ representing various inter and intra-band interactions. 
Below we
show how three characteristic length scales are defined in this two component model
(two associated with density variations and
the London magnetic field penetration length).

The only vortex solutions of the model (\ref{gl}) which have
finite energy per unit length are the integer $N$-flux quantum
vortices which have the following phase windings along a contour
$l$  around the vortex core: $\oint_l \nabla \theta_1= 2\pi N,
\oint_l \nabla \theta_2= 2\pi N$ which can be denoted as (N,N).
Vortices with differing phase windings (N,M) carry a fractional
multiple of the
 magnetic
flux quantum and have energy divergent with the system size
\cite{frac}, which under usual conditions makes them irrelevant 
for the physics of magnetic response of a bulk system.

In what follows  we investigate only the
integer flux vortex solutions which
are the energetically cheapest objects to produce by means of an
external field in a bulk superconductor. Note that since this object is essentially
a  bound state of two vortices, it in general will have two different co-centered cores.

\section{Coherence lengths and intervortex forces at long range in multi-band superconductors}
\label{asymp}

In this section we give criterion for  attractive or repulsive 
 force between
well separated vortices in system (\ref{gl}) 
and condition for non-monotonic inter vortex forces
following \cite{Babaev.Speight:05,johan1,johan2}.
We show how it can be determined  by
analyzing $F_p$ when $\nu=0$ and how three fundamental length scales are
defined in the model (\ref{gl}). If the model has mixed gradient terms they can
be either treated as in Ref. \cite{johan2} or eliminated by a linear transformation.
By gauge invariance, $F_p$ may depend only on $|\psi_1|$, $|\psi_2|$ and
$\delta=\theta_1-\theta_2$. We consider the regime  { when $F_p$} has a global minimum at some point
other than the one with $|\psi_i|=0$,  {namely at} 
$(|\psi_1|,|\psi_2|,\delta)=(u_1,u_2,0)$ where $u_1>0$ and $u_2\geq 0$ (for discussion of phase-separated regimes see
\cite{Garaud.Babaev:15}).
Then the model has a trivial solution,
$\psi_1=u_1$, $\psi_2=u_2$, $A=0$, (i.e. the ground state).
Here we are interested in models that support
axially-symmetric single-vortex solutions of the form
\begin{align}
 \psi_i&=f_i(r)e^{i\theta}\,,&
(A_1,A_2)&=\frac{a(r)}{r}(-\sin\theta,\cos\theta)\label{ansatz}
\end{align}
where $f_1,f_2,a$ are real profile functions with boundary behavior
$f_i(0)=a(0)=0$, $f_i(\infty)=u_i$, $a(\infty)=-1/e$. No explicit
expressions for $f_i,a$ are known, but, by analyzing the  system
of differential equations they
satisfy, one can construct asymptotic expansions for them at large $r$,
see 
\cite{johan1,johan2}.

At large $r$ from  the vortex in the model
(\ref{gl})  the system
recovers (up to exponentially small
corrections) the ground state. In fact, the long-range field
behavior of a vortex solution can be identified
with a solution of the linearization of the model about the ground state,
in the presence of appropriate point sources at the vortex positions. This idea
is explained in detail for single component GL theory in \cite{spe}.
A common feature of topological solitons (vortices being
a particular example) is that the forces they
exert on one another coincide asymptotically (at large separation) with
those between the corresponding point-like perturbations (point sources) interacting via the linearized
field theory \cite{Manton.Sutcliffe}. For (\ref{gl}), the linearization has one vector
($A$) and 3 real
scalar ($\eps_1=|\psi_1|-u_1$, $\eps_2=|\psi_2|-u_2$ and $\delta$)
degrees of freedom.
The isolated vortex solutions have, by definition within the
ansatz  we use, $\delta\equiv 0$ everywhere.
Hence have no source for $\delta$, so we can set $\delta=0$ in the
linearization, which becomes
\be
F_{lin}=\frac12|\nabla\eps_1|^2+\frac12|\nabla\eps_2|^2+
\frac12\left(\begin{array}{c}\eps_1\\ \eps_2\end{array}\right)\cdot
\hh\left(\begin{array}{c}\eps_1\\ \eps_2\end{array}\right)
\label{gllin}
+\frac12(\cd_1A_2-\cd_2A_1)^2
+\frac12e^2(u_1^2+u_2^2)|A|^2.
\ee
Here, $\hh$ is the Hessian matrix of $F_p(|\psi_1|,|\psi_2|,0)$
about $(u_1,u_2)$, that is,
\beq \label{Eq:Hessian}
\hh_{ij}=\left.\frac{\cd^2F_p}{\cd|\psi_i|\cd|\psi_j|}\right|_{(u_1,u_2,0)}.
\eeq
Note that, in $F_{lin}$, the vector potential field $A$ decouples from the scalar fields $\psi_i$.
This mode
 mediates a repulsive force between vortices (originating in current-current and magnetic interaction)
with decay length which is the London's magnetic field penetration length  $\lambda=1/\mu_A$ where $\mu_A$
is the mass of the field, that is,
\beq
\mu_A=e\sqrt{u_1^2+u_2^2}.
\eeq

By contrast, the scalar fields $\eps_1,\eps_2$ are, in general, coupled (i.e.\
 the symmetric matrix $\hh$ has off-diagonal terms). To remove the cross-terms
one should find a proper linear combination of the fields that correspond to normal modes of the system.
To this end we  make a linear redefinition of fields, expanding
$(\eps_1,\eps_2)^T$ with respect to the orthonormal basis for $\R^2$
formed by the eigenvectors $v_1,v_2$ of $\hh$,
\beq
(\eps_1,\eps_2)^T=\chi_1 v_1+\chi_2 v_2.
\eeq
 The corresponding
eigenvalues $\mu_1^2,\mu_2^2$ are necessarily real (since $\hh$ is
symmetric) and positive (since $(u_1,u_2)$ is a minimum of $F_p$), and hence
\be
F_{lin}=\frac12\sum_{a=1}^2\left(|\nabla\chi_i|^2+\mu_i^2\chi_i^2\right)
+\frac12(\cd_1A_2-\cd_2A_1)^2
+\frac12e(u_1^2+u_2^2)|A|^2.
\ee
The scalar fields $\chi_1,\chi_2$  describe linear combination 
of the original density fields. The new fields  recover ground state values at
different characteristic length scales. The characteristic length scales are nothing but coherence lengths which
are given by inverse of $\mu_i$ 
\begin{align}
\xi_1&\equiv 1/\mu_1 \,, &
\xi_2& \equiv 1/\mu_2 
\end{align}
 respectively {\it (here and below we absorb  the factor $1/\sqrt{2}$ in the definition of coherence length)}.
 Each of these field defines a vortex
core of some characteristic size that mediate an attractive force between
vortices at long range. 
 In terms of the normal-mode fields $\chi_1,\chi_2$ and $A$, the composite point source
which must be introduced into $F_{lin}$ to produce field configurations
identical to those of  vortex asymptotics is
\begin{align}
\kappa_1&=q_1\delta(x)\,, &	
\kappa_2&=q_2\delta(x)\,, &	
{\bf j}&=m(\cd_2,-\cd_1)\delta(x)\label{cps} \,,
\end{align}
where $\kappa_1$ is the source for $\chi_1$, $\kappa_2$ the source of $\chi_2$, ${\bf j}$ the source for ${\bf A}$, $\delta(x)$ denotes the two
dimensional Dirac delta function and $q_1,q_2$ and $m$ are unknown real constants which can, in principle, be determined numerically by
a careful analysis of the vortex asymptotics. Physically,  a vortex, as seen
from a long distance  can be thought of as a point particle carrying two different
types of scalar monopole charge, $q_1,q_2$, inducing fields of mass $\mu_1,\mu_2$ respectively, and a magnetic dipole moment $m$ oriented orthogonal to the
$x_1x_2$ plane, inducing a massive vector field of mass $\mu_A\equiv   \lambda^{-1}$. 
The interaction energy experienced by a pair of point particles carrying these sources, held distance $r$ apart,
is easily computed in linear field theory. For example, two scalar monopoles of charge $q$ inducing fields of mass $\mu$ held at positions
${\bf y}$ and $\tilde{\bf y}$ in $\R^2$ experience interaction energy
\be
E_{int}=-\int_{\R^2} \kappa\tilde\chi=-\int_{\R^2}q\delta({\bf x}-{\bf y})\frac{q}{2\pi}K_0(\mu |{\bf y}-\tilde{\bf y}|)
=-\frac{q^2}{2\pi}K_0(\mu|{\bf y}-\tilde{\bf y}|)
\ee
where $\kappa$ is the source for the monopole at ${\bf y}$, $\tilde\chi$ is the scalar field induced by the monopole at $\tilde{\bf y}$ \cite{spe} and
$K_0$ denotes the modified Bessel's function of the second kind. The interaction energy for a pair of magnetic dipoles may be computed similarly. In the case
of our two component GL model, the total long-range inter-vortex interaction energy has three terms, corresponding to the three sources in the composite point source (\ref{cps}),
and turns out to be
\beq
E_{int}=\frac{m^2}{2\pi}K_0(\mu_A r)-\frac{q_1^2}{2\pi}K_0(\mu_1 r)-\frac{q_2^2}{2\pi}K_0(\mu_2 r).
\label{eint}
\eeq
Note that, the first term in this formula which
originates in magnetic and current-current
interaction is repulsive, while the other two as associated
with core-core interactions of two kinds of co-centered cores are attractive. 
The linearized theory does not contain information about the  prefactors $q_1,q_2$ and $m$. However they can be determined
numerically from the full nonlinear GL theory.
At very large $r$, $E_{int}(r)$ is dominated by whichever term
corresponds to the smallest of the three masses, $\mu_A$, $\mu_1$, $\mu_2$, so to determine whether vortices attract at long range, it is enough to compute
just these masses. The generalization to the case with larger number of components is straightforward:
additional coherence lengths give additional contributions to attractive interaction in the form
$-\frac{q_i^2}{2\pi}K_0(\mu_i r)$.  Generalizations to multiple repulsive length scales in layered systems or caused by stray fields were discussed in \cite{varney2013hierarchical} .  In thin films intervortex interaction acquires also $1/r$ repulsion at long ranges due to the magnetic field outside the sample, similarly to single-component case  \cite{pearl}.

Consider the case where  long-range interaction
is attractive due to $\xi_1$ being the largest length scale of the problem.
 The criterion  for short-range repulsive interaction is thermodynamic stability  of vortices which is equivalent to the condition that the system has solution with 
negative free energy interfaces in external fields \cite{Babaev.Speight:05,johan1,johan2}. Indeed
when the interface energy is always positive the system exhibits type-1 behavior: i.e. tends to form a 
single    vortex with high finding number. If there are interfaces with negative energy in the external field, the system tends to
maximize these interfaces. In the type-1.5 regime the system  forms vortex
clusters, where it maximizes number of vortex cores inside the vortex clusters.
A the same time the system minimizes the
interface of the cluster itself (that costs positive energy).

To summarize, the nature of intervortex forces at large separation in the model 
under consideration, can
be determined purely by analyzing $F_p$: one finds the ground state $(u_1,u_2)$
and  the Hessian $\hh$ of $F_p$
about $(u_1,u_2)$. From this one computes the mass of the
vector field $A$, $\mu_A=e\sqrt{u_1^2+u_2^2}$ (i.e. the inverse of  the magnetic field penetration length), and the masses
$\mu_1,\mu_2$ of the scalar normal modes 
(i.e. the inverses of  the coherence lengths). 
These masses being the square roots of
the eigenvalues of $\hh$. If either (or both) of $\mu_1,\mu_2$ are less than
$\mu_A$, then the dominant interaction at long range is attractive (i.e.
vortex core extends beyond the area where magnetic field is localized), while
if $\mu_A$ is less than both $\mu_1$ and $\mu_2$, the dominant interaction
at long range is repulsive.
The special feature of the two-component model is that the vortices
where core extends beyond the magnetic field penetration length
are thermodynamically stable in a range of parameters and moreover one can have
a repulsive force between the vortices at shorter distances 
where the system has thermodynamically stable vortex solutions \cite{Babaev.Speight:05,johan1,johan2}.
It is important to stress that length scales $\mu_1^{-1},
\mu_2^{-1}$  are not directly
associated with the individual condensates $\psi_1$, $\psi_2$.
Rather they are associated with the normal modes $\chi_1,\chi_2$, defined as \cite{johan1,johan2}
\begin{equation}
 \chi_1=(|\psi_1|-u_1)\cos\Theta-(|\psi_2|-u_2)\sin\Theta\,, ~~~
\chi_2=-(|\psi_1|-u_1)\sin\Theta-(|\psi_2|-u_2)\cos\Theta\,.
\end{equation}
These may be thought of as
rotated (in field space) versions of $\eps_1=|\psi_1|-u_1$,
$\eps_2=|\psi_2|-u_2$. The {\em mixing angle}, that is, the angle between
the $\chi$ and $\eps$ axes, is $\Theta$, where the eigenvector
$v_1$ of $\hh$ is $(\cos\Theta,\sin\Theta)^T$. This, again, can be determined
directly from $\hh$.

Note also that the shorter of the length scales  $\mu_1^{-1},
\mu_2^{-1}$, although being a fundamental length scale of the theory,
can be masked in a density profile of a vortex solution by nonlinear effects.
This, for example 
certainly happens if   $\mu_1^{-1} \ll \mu_A \equiv \lambda^{-1}$ (see short discussion in Ref. \cite{johan1}). 
Also note that in general the minimum of the interaction potential will not be located  at the London penetration length, because it in general will be also affected by nonlinearities.

From the discussion above it  follows that in general one cannot drop the subdominant component based on 
comparison of the ground state values of the amplitudes of $|\psi_i |$ in the GL expansion.
 Namely, the long-range interaction can be determined by a mode
with smaller amplitude.  The formal justification of the multiband GL expansion can be found in \cite{Silaev.Babaev:12}.

\subsection{Example: a  superconductor with a passive band}

To illustrate the analysis of the coherence lengths presented above, we consider the simple case of a
two band superconductor where one of the bands is passive, that is, with a
potential of the form
\beq
F_p=-\alpha_1|\psi_1|^2+\frac{\beta_1}{2}|\psi_1|^2+\alpha_2|\psi_2|^2
-\gamma(\psi_1 \psi_2^*+ \psi_1^*\psi_2)
\eeq
where $\alpha_j,\beta_1,\gamma$ are positive constants. Then $F_p$ is 
minimized when $\psi_1$ and $\psi_2$ have equal phase, and have moduli
\beq
|\psi_1|=u_1=\sqrt{\frac{\alpha_1}{\beta_1}\left(1+\frac{\gamma^2}{\alpha_1\alpha_2}\right)},\qquad
|\psi_2|=u_2=\frac{\gamma}{\alpha_2}u_1.
\eeq
The mass of the vector field $A$ is 
\beq
\mu_A=e\sqrt{u_1^2+u_2^2}=eu_1\sqrt{1+\frac{\gamma^2}{\alpha_2^2}}.
\eeq
The Hessian matrix of $F_p$ about $(u_1,u_2)$ is
\beq
\hh=\left(\begin{array}{cc}4\alpha_1+\frac{6\gamma^2}{\alpha_2}&-2\gamma\\
-2\gamma&2\alpha_2\end{array}\right).
\eeq
It is straightforward to compute explicit expressions for the eigenvalues
$\mu_1^2,\mu_2^2$ of this matrix. The 
power series expansion in $\gamma$ reveals that
\beq
\mu_1=2\sqrt{\alpha_1}+O(\gamma^2),\qquad
\mu_2=\sqrt{2\alpha_2}+O(\gamma^2).
\eeq
Similarly, the normalized eigenvector associated with eigenvalue $\mu_1^2$ is
\beq
v_1=\left(\begin{array}{c}1\\-(2\alpha_1-\alpha_2)^{-1}\gamma\end{array}\right)
+O(\gamma^2)\eeq
so the normal modes of fluctuation about the ground state are rotated through
a mixing angle 
\beq \label{Eq:MixingAngle0}
\Theta=-(2\alpha_1-\alpha_2)^{-1}\gamma+O(\gamma^2).
\eeq
In comparison with the uncoupled model ($\gamma=0$) then, we see that, for
small coupling $\gamma$ the length scales $\lambda=1/\mu_A,\xi_1=1/\mu_1,\xi_2=1/\mu_2$ are
unchanged to leading order, but the normal modes with which $1/\mu_1,1/\mu_2$
are associated are mixed to leading order. In particular, there are
large regions of parameter space where $\mu_2<\mu_A<\mu_1$, so that
vortices attract at long range, even though the active band, $\psi_1$, is
naively ``type-2' (that is, $\beta_1>e^2/4$).

\section{Critical coupling (Bogomolnyi point)}\label{Bogom}
Although it is not related to the topic of this paper, in this section we briefly
review Bogomolnyi point physics.
In single-component superconductors, the type-1 and type-2 regimes are
separated by a Bogomolnyi point $\kappa_c=1$ (note, again  that above we absorbed the factor $1/\sqrt{2}$ into the definition of coherence length, for this reason the critical coupling is different from $1/\sqrt{2}$). At that point 
vortices do not interact, the free energy of normal-to-superconductor interfaces is zero and we have $H_{c1}=H_{c2}=H_c$\cite{Bogomol,bogomol1976stability,saint1969type,Manton.Sutcliffe}.
This regime is referred to as the ``critical point"  because of the saturation of Bogomolnyi  inequality  \cite{Bogomol,bogomol1976stability,saint1969type,Manton.Sutcliffe,shifman2012advanced,ssm}.
The necessary but not sufficient conditions for a critical point is lack of intervortex
forces at long range within the linear approximation. To that end   all   modes excited in a vortex solution, must have equal masses $\mu_i$. From eq. (\ref{eint}) it is obvious that   for a multicomponent superconductor it requires
a fine-tuning and in general type-1 and type-2 regimes are not separated by a 
critical  point.  
Furthermore from the section on microscopic theory below it is clear that 
in general  $\mu_1$ and $\mu_2$ (as  functions of system's parameters and temperature) do not cross
but form an avoided crossing. Thus in the two-component case the Bogomolnyi critical point is a zero-measure parameter set which 
requires special symmetry of the model. 
Such a fine tuning for a composite vortex can be achieved in $U(1)\times U(1)$ system
with a potential that is symmetric with respect to both components

\beq
F_p=-\alpha |\psi_1|^2+\frac{\beta}{2}|\psi_1|^2-\alpha |\psi_2|^2+\frac{\beta}{2}|\psi_2|^2
\eeq 

For a standard form of gradient terms this potential
gives equal coherence lengths. The Bogomolnyi 
point is realised when $\xi_1=\xi_2=\lambda$, just like in single-component
system vortices do not interact in this regime.

The lack of interaction between vortices at Bogomolnyi point originates in 
exact cancelation of electromagnetic and core-core interaction.
However  some weak interaction indeed appear beyond the Ginzburg-Landau theory
e.g. by non-locality effects, as was studies in detail
single-component models  \cite{jacobs,leung1,eilenberger,PhysRevB.4.3029}. 
This opens up a narrow window in the parameter space
 near $\kappa\approx 1$ where the interaction, as was suggested by Eilenberger
 and studies in detail by Jacobs et al is non-monotonic 
 and in general   depends
 on microscopic detail \cite{jacobs,leung1,eilenberger}.
 The simplest approach that was used  to investigate the $\kappa\approx 1$
 regime was to carry single-component GL expansion to 
 next to leading order in $\tau=1-T/T_c $  and gradients \cite{ovchinnikov1999generalized,Ovchinnikov2013}.
However to estimate width of the region, full microscopic theory is need. This was
studied in \cite{jacobs,leung1,eilenberger,PhysRevB.4.3029}.

 In two-band superconductors  similar ``near-Bogomolnyi" regime can
 be realised  in  the regimes where  there are two-bands but there is no second coherence length
 and system is characterized by a single parameter $\kappa\approx 1$ [such situations appear e.g. when there is strong interband coupling \cite{Silaev.Babaev:11,Silaev.Babaev:12}].  Two-band microscopic theory 
 confirmed that 
the intervortex interaction that play role in ``near-Bogomolnyi" regime in single-component model,
are important only in the narrow window of $\kappa \approx 1$
 in  two-band materials well characterized by single $\kappa$ and is negligible otherwise. The multi-band materials where this physics should be
 relevant for intervortex interaction should have $\kappa\approx 1$ and $H_{c1}\approx H_{c2}$.  These microscopic corrections  cannot give observable intervortex attraction of
 materials like  the introduction $MgB_2$ and $Sr_2RuO_4$.

In the type-1.5 regime the dominant intervortex forces have different origin and  the effects responsible
for intervortex forces in the ``near-Bogomolnyi" point (non locality etc) are not important.
For a comparative study of this physics in multiband materials, full microscopic theory is required that accounts both for
multiple coherence lengths and nonlocal effects. It was presented
in \cite{Silaev.Babaev:11}, certain aspects of it are discussed in the section below.

 { \section{Microscopic theory of type-1.5 superconductivity in $U(1)$ multiband case}}
 \label{microscopic}

 In this section we briefly outline microscopic {theory of type-1.5 superconductivity} in the particular case
of clean multi-band superconductors that break only $U(1)$ symmetry.
In this case existence of multiple coherence lengths does not follow from symmetry and  
has to be justified.  A reader who is interested in more general cases of higher symmetry
breaking as well the general properties of the type-1.5 state can skip this discussion and proceed directly to the next section. Existence of multiple superconducting bands is not 
a necessary condition for appearance of multiple coherence lengths  \cite{Silaev.Babaev:11}. 
The appearance of multiple  coherence lengths  and  type-1.5 regime  in multiband-band superconductors was described using microscopic theory at all temperatures,
without relying on GL expansions in \cite{Silaev.Babaev:11}. We refer a reader, interested in a full  microscopic theory
that does not rely on GL expansion to that work, while here we mainly focus on microscopic justification of GL expansion.

 As discussed above, in multi-band systems, in general multi-component GL expansions are not based on symmetry (indeed it is a particular example when  a soft explicit breaking
 of a higher symmetry does not necessarily eliminate classical-field-theoretic description of
 non-critical modes). Therefore obviously it cannot be obtained as an expansion in  a single small parameter $\tau=1-T/T_c$. 
Instead such expansions are justified when the system has multiple  small parameters
which are not symmetry-related. 
In the simplest case these are  multiple  small gaps in different bands, small gradients, and small interband coupling constants.
A single-parameter-$\tau$ expansion emerges as a single-component reduction of the model in the $\tau\to 0$ limit for a system that breaks only $U(1)$ symmetry \cite{Silaev.Babaev:12}. 
 
  In this section we focus on  two-band case  and consider the microscopic origin of two-component GL model 
 (TCGL) described by the following free energy density:
\begin{eqnarray} 
 F &=&  \sum_{j=1,2}\left(a_j | \Delta_j |^2
 + \frac{b_j}{2}| \Delta_j|^4 + K_j |\bm D\Delta_j |^2\right)   \nonumber\\
 & -&   \gamma \left( \Delta_1\Delta_2^*+\Delta_2\Delta_1^*\right)+\frac{ B ^2}{8\pi } 
\label{Gibbs}
\end{eqnarray}
 where ${\bm D}={\bm \nabla}  + i {\bm A}$, ${\bm A}$ and ${\bm B}$ are the vector potential 
 and magnetic field and  $\Delta_{1,2}$ are the gap functions in two different bands. 

\subsection{ Microscopic model for $U(1)$ two-band system.}
\label{App:GLexpansion}


To verify applicability of TCGL theory we consider the microscopic model of a clean superconductor with two
overlapping bands at the Fermi level \cite{Silaev.Babaev:11,Silaev.Babaev:12}. Within
quasiclassical approximation the band parameters characterizing
the two different cylindrical sheets of the Fermi surface are the
Fermi velocities $V_{Fj}$ and the partial densities of states
(DOS) $\nu_j$, labelled by the band index $j=1,2$.

It is convenient to normalize the energies to the critical
temperature $T_c$ and length to $r_0= \hbar V_{F1}/T_c$. The
vector potential is normalized by $\phi_0 /(2\pi r_0)$, the
current density normalized by $c\phi_0/(8\pi^2 r_0^3)$ and
therefore the magnetic field is measured in units
 $\phi_0 /(2\pi r^2_0)$ where $\phi_0=\pi\hbar c/e$ is the magnetic flux quantum.
In these units the Eilenberger equations for quasiclassical
propagators take the form
\begin{align}\label{Eq:EilenbergerF}
&v_{Fj}{\bm n_p}{\bm D} f_j +
 2\omega_n f_j - 2 \Delta_j g_j=0, \\ \nonumber
 &v_{Fj}{\bm n_p}{\bm D}^* f^+_j -
 2\omega_n f^+_j + 2\Delta^*_j g_j=0.
 \end{align}
 Here $v_{Fj}=V_{Fj}/V_{F1}$, $\omega_n=(2n+1)\pi T$ are Matsubara frequencies,
   the vector ${\bm n_p}=(\cos\theta_p,\sin\theta_p)$
  parametrizes the position on 2D cylindrical
 Fermi surfaces. The quasiclassical Green's functions in each band obey
 normalization condition $g_j^2+f_jf_j^+=1$.

 The self-consistency equation for the gaps is
\begin{equation}\label{Eq:SelfConsistentGap}
 \Delta_i=T \sum_{n=0}^{N_d} \int_0^{2\pi}
 \lambda_{ij} f_j d\theta_p.
\end{equation}
 The coupling matrix $\lambda_{ij}$ satisfies the symmetry relations
 $n_1\lambda_{12}=n_2\lambda_{21}$ where $n_i$ are the
 partial densities of states normalized so that $n_1+n_2=1$.
 The vector potential satisfies the Maxwell equation
 $\nabla\times\nabla\times {\bm A} = {\bm j}$ where the current is
 \begin{equation}\label{Eq:SelfConsistentCurrent}
  {\bm j}= -T\sum_{j=1,2} \sigma_j\sum_{n=0}^{N_d}
 Im\int_0^{2\pi}  {\bm n_p} g_j d\theta_p.
\end{equation}
 The parameters $\sigma_j$ are given by
 $ \sigma_j=4\pi\rho n_j v_{Fj}$ and
 $ \rho=(2e/c)^2 (r_0 V_{F1})^2\nu_0 $.
 
\begin{figure}
\centerline{\includegraphics[width=0.60\linewidth]{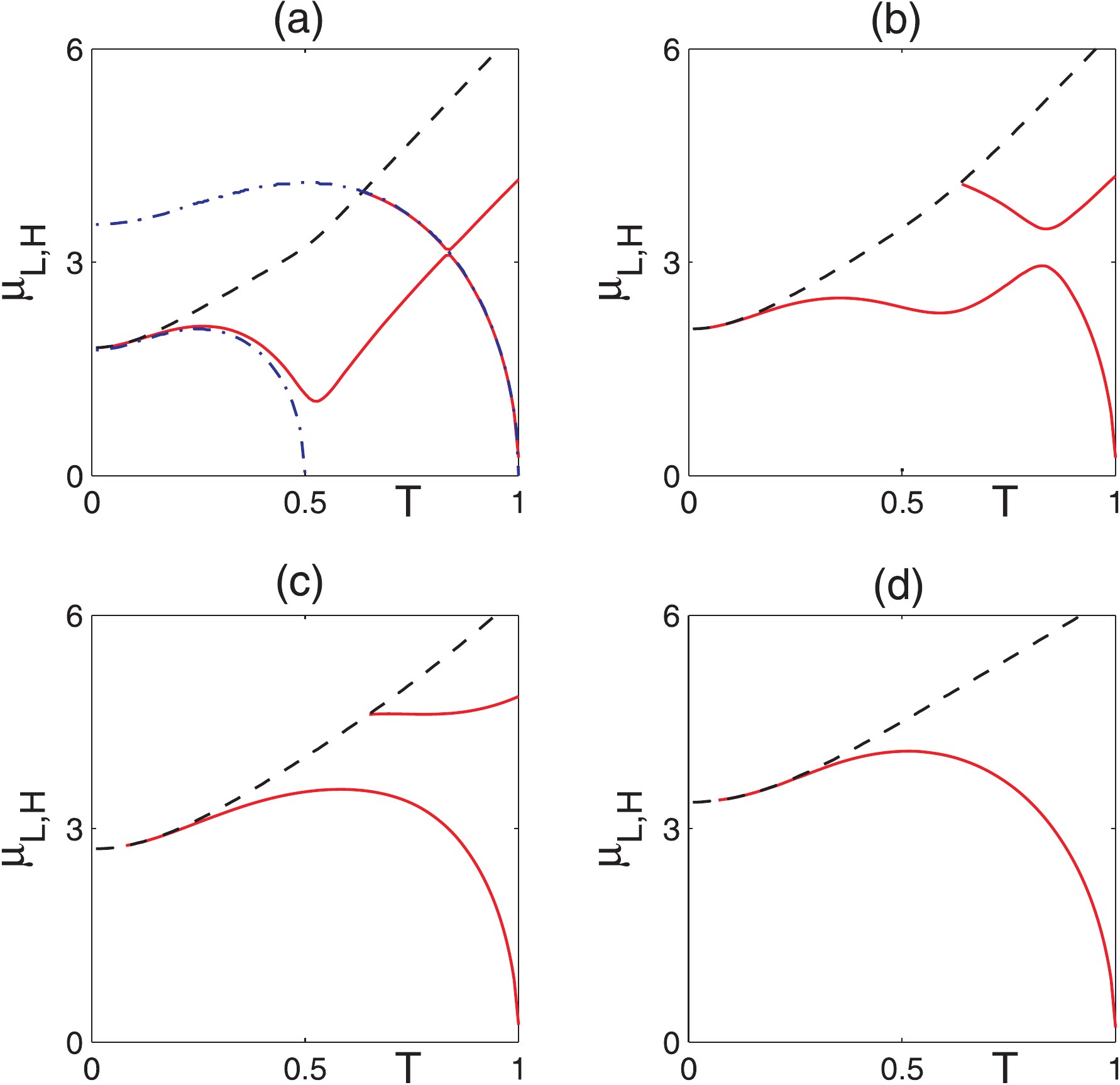}}
\caption{\label{Fig:SequenceModes} Calculated in \cite{Silaev.Babaev:11} the inverse of coherence lengths (or masses) $\mu_{L}$ and $\mu_{H}$
(red solid lines) of the composite gap function fields for the
different values of interband Josephson coupling $\lambda_J$ and
$v_{Fk}=1$. 
 By black dash-dotted lines the branch cuts are
shown.
The  coupling
constants are $\lambda_{11}=0.25$, $\lambda_{22}=0.213$ 
and $\lambda_J=0.0005;\;0.0025;\;0.025;\;0.213$ for plots (a-d)
correspondingly.
  In (a) the blue dash-dotted lines correspond to $\lambda_J=0$
showing two masses going to zero at two different temperatures. 
This corresponds to $U(1)\times U(1)$ theory with two
 coherence lengths that diverge at different temperatures. For  $\lambda_J\neq 0$ only $\mu_L$ goes to zero at
 $T_c$: this is  a consequence of the fact that
Josephson coupling breaks the symmetry explicitly down to single $U(1)$ and thus there can be only one diverging coherence length.
As one can see from the panels (a)-(c) the $U(1)$ system nonetheless has two well-defined coherence lengths in these regimes, that implies that there are two characteristic length scales in spatial variations of two different linear combinations of fields.
When symmetry breaking is weak there is a local maximum of the larger   coherence lengths (local minimum of $\mu_L$)  at lower temperature (panels (a) and (b)).  }
\end{figure}

 { First we can use the microscopic model formulated above to check if it is consistent with the basic 
 feature of the multicomponent GL theory, namely the existence of several distinct coherence lengths.
 For that purpose we calculate asymptotics of gap
 functions $|\Delta_{1,2}|(r)$ linearizing the
 systems of Eilenberger equations (\ref{Eq:EilenbergerF}) together with the
 self-consistency equation (\ref{Eq:SelfConsistentGap}).
 In momentum representation this linear system
 has a discrete and continuous spectrum of imaginary eigenvalues    
 which determine the inverse decay scales of the gap functions 
 deviations from the ground state\cite{Silaev.Babaev:11}. 
 The discrete eigenvalues correspond to the 
  masses of gap functions fields (i.e. the inverse of coherence lengths 
  associated with the linear combinations of the fields as in the previous sections). 
  Besides that there is a continuous spectrum which contains all scales 
  lying on the branch cut going along the imaginary axis starting 
  at $\min_k 2\left(\sqrt{|\Delta_k|^2+(\pi T)^2 }/v_{Fk}\right)$.

 The examples from ref. \cite{Silaev.Babaev:11} of the temperature dependencies
 of the masses $\mu_{L,H}(T)$ are shown in the Fig.\ref{Fig:SequenceModes}.
 The evolution of the masses $\mu_{L,H}$ is shown in the sequence
 of plots Fig.\ref{Fig:SequenceModes}(a)-(d)  for $\lambda_J$
 increasing from the small values $\lambda_J\ll \lambda_{11},\lambda_{22}$ to the values comparable to intraband
 coupling $\lambda_J\sim \lambda_{11},\lambda_{22}$. 
  The regime with two massive modes is exactly
 the same as the given by a two-component GL theory.
 In this model, at very low temperatures there exists only one massive mode $\mu_{L}(T)$
 lying below the branch cut.
 As the interband coupling is increased above some critical value  the mass
 $\mu_{H}(T)$ disappears at all temperatures. In such a regime the asymptotic is determined by a single coherence length and a branch cut with a continuous spectrum of scales.
 The branch cut contribution is essentially a non-local 
 effect which is not captured by GL theory. Therefore one
 can expect growing discrepancies between effective GL solution and
 the result of microscopic theory at low temperatures.  }

\begin{figure}
\centerline{\includegraphics[width=0.60\linewidth]{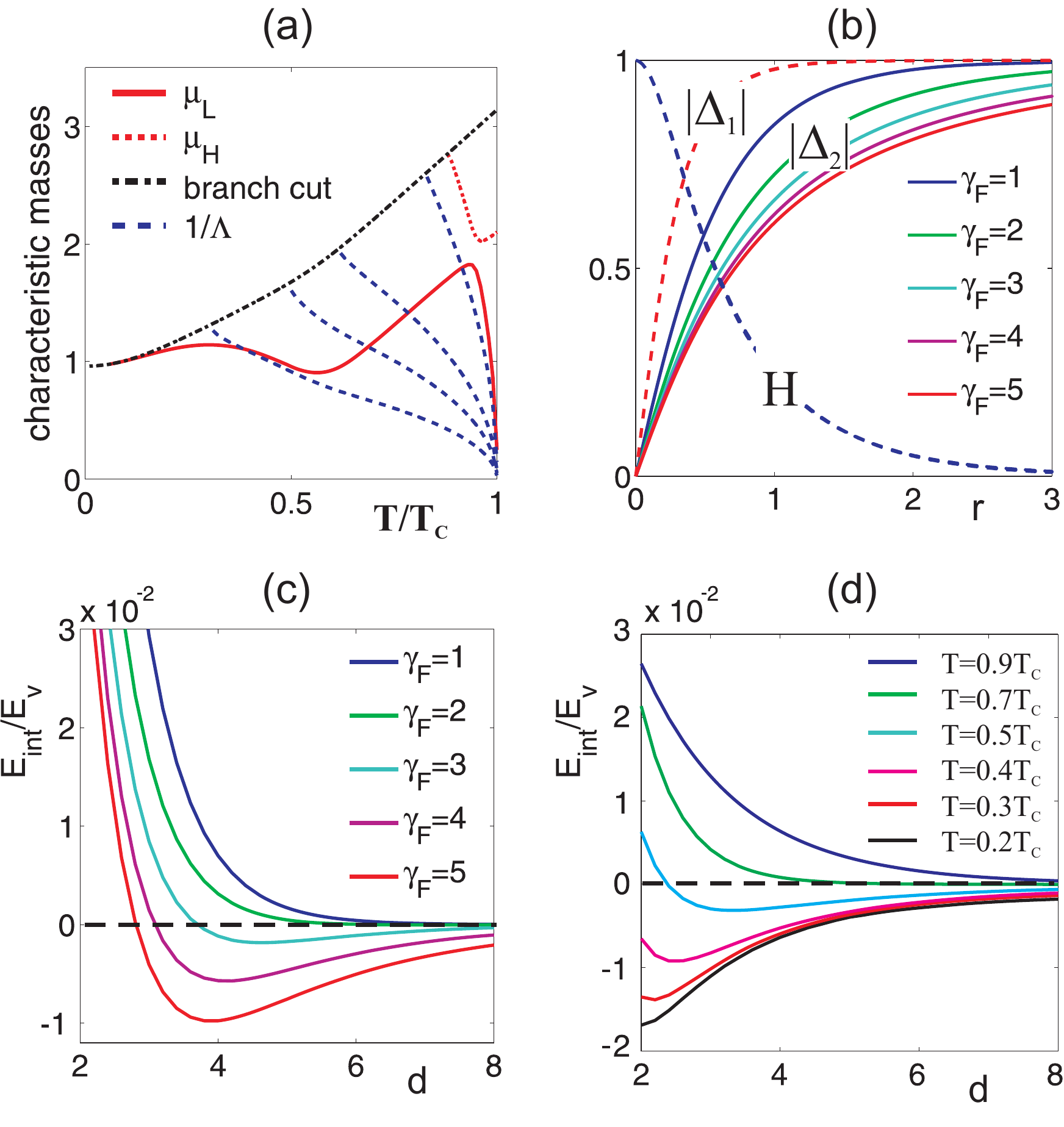}}
\caption{\label{Fig:VortexStructure15} Calculated in ref. \cite{Silaev.Babaev:11} (a) the inverse of coherence lengths (or masses) $\mu_{L}$ and
$\mu_{H}$ (red solid and dotted lines) of the composite gap
function fields and inverse London penetration (blue dashed lines)
for the different values of $\Lambda \mu_L
(T_c)/\sqrt{2}=1;2;3;5$. { Scales above the black dash-dotted line are contained in a branch-cut contribution. }
 (b) Distributions of magnetic field $H(r)/H(r=0)$, gap functions
$|\Delta_1|(r)/\Delta_{10}$ (dashed lines) and
$|\Delta_2|(r)/\Delta_{20}$ (solid lines) for the coupling
parameters $\lambda_{11}=0.25$, $\lambda_{22}=0.213$ and
$\lambda_{21}=0.0025$ and different values of the band parameter
$\gamma_F=1;2;3;4;5$. { (c,d) The   interaction energy $E_{int}$ between two
vortices normalized to the single vortex energy $E_v$ as function of the
intervortex distance $d$. In panel (c) the pairing constants are the same as in (d) and $T=0.6 T_c$. 
In panel (d) $\lambda_{ii}$ are the same as in (c) but $\lambda_{21}=0.00125$, and $\gamma_F=1$. }
 The panel (d) shows  that when, at lower temperature, one 
of the coherence lengths becomes larger than the magnetic
field penetration length, the system falls into type-1.5 regime.
The strength on the attraction grows as temperature becomes lower.
The inter vortex distance corresponding to minimum of interaction
potential is strongly temperature dependent
even below $0.5 T_c$. It diminishes when temperature is decreased
which should result in vortex cluster shrinkage with decreasing temperature. 
 }
\end{figure}

 { 
 Besides justifying the predictions of phenomenological two-component GL theory \cite{Silaev.Babaev:12} the
microscopic formalism developed in Ref.(\cite{Silaev.Babaev:11}) allows to
describe type-1.5 superconductivity beyond the validity of GL
models.  As shown on Fig.\ref{Fig:VortexStructure15}a
the function $\mu_L(T)$ is {\it non-monotonic} at low
temperatures. Therefore for a certain range of parameters in contrast with the
physics of singe-band superconductors the product of London
penetration depth $\Lambda$ and $\mu_L$ has a strong and
nonmonotonic temperature dependence 
 making intervortex attraction possible if $\Lambda \mu_L<1$. 

To demonstrate the type-1.5 superconductivity i.e. large-scale
attraction and small-scale repulsion of vortices which originates
from disparity of two coherence lengths,
 the inter-vortex interaction energy $E_{int}$ was  calculated in \cite{Silaev.Babaev:11}.
 In Fig.\ref{Fig:VortexStructure15}(c,d) $E_{int}$ (normalized to the single
vortex energy $E_v$) is shown as a function of the distance between
two vortices $d$. The plots on
Fig.\ref{Fig:VortexStructure15}(c) clearly demonstrate the
emergence of type-1.5 behavior  when the parameter $\gamma_F=v_{F1}/v_{F2}$, which characterizes
the disparity in band characteristics  is increased. 

 {  The type-1.5 regime manifested in
 the appearance a non-monotonic behavior of $E_{int}(d)$
 when one of the coherence length becomes larger than the magnetic field penetration length.
 The  Fig.\ref{Fig:VortexStructure15}(d) shows $E_{int}$ }
for a $U(1)$ two band superconductor that is type-2 near $T_c$
and becomes type-1.5  near the  temperature where one band
crosses over from interband-proximity-effect-induced to intrinsic superconductivity.
The long-range attractive forces in type-1.5 regime are similar to
the long-range forces in type-1 superconductors, while short-range forces are similar to those
in type-2 superconductors. The physical origin and form of these interactions are obviously  principally different from the discussed in section \ref{Bogom} microscopic-physics and non-locality-dominated 
intervortex forces in superconductors near Bogomolnyi point.}

\subsection{ Microscopic Ginzburg-Landau theory  for $U(1)$ two-band system.}

Here we briefly outline the derivation of   TCGL functional (\ref{Gibbs}) from the
microscopic equations  following \cite{Silaev.Babaev:12}.  
First  we
find the solutions of Eilenberger Eqs.(\ref{Eq:EilenbergerF}) in
the form of the expansion by the amplitudes of gap functions 
$|\Delta_{1,2}|$ and their gradients $|({\bm n_p \bm D})
\Delta_{1,2}|$. Then these solutions are substituted to the
self-consistency Eq.(\ref{Eq:SelfConsistentGap}). Using this
procedure we find the solutions of Eqs.(\ref{Eq:EilenbergerF}) in
the form:
\begin{equation}\label{Eq:2OrderExpansion}
f_j=\frac{\Delta_j}{\omega_n}-\frac{|\Delta_j|^2\Delta_j}{2\omega_n^3}-\frac{v_{Fj}}{2\omega_n^2}
({\bm n_p\bm D}) \Delta_j+\frac{v_{Fj}^2}{4\omega_n^3} ({\bm n_p \bm D}) ({\bm n_p\bm D}) \Delta_j.
\end{equation}
and $f^+_j({\bm n_p})=f^*_j(-{\bm n_p})$. Note that this GL expansion is
based on neglecting the higher-order terms in powers of $|\Delta|$
and $|({\bm n_p\bm D}) \Delta|$. Indeed this approximation
naturally fails in a number of cases. 
The regimes when it can be justified were determined in the work \cite{Silaev.Babaev:12}
 by a direct comparison to the full microscopic model. 
  Let us determine
microscopic coefficients in the GL expansion.
Substituting to the self-consistency
Eqs.(\ref{Eq:SelfConsistentGap}) and integrating by $\theta_p$ we
obtain
  \begin{eqnarray}\label{}
  \Delta_1=(\lambda_{11} \Delta_1+\lambda_{12} \Delta_2)  G +  (\lambda_{11} GL_1+\lambda_{12} GL_2) \\
  \Delta_2=(\lambda_{21} \Delta_1+\lambda_{22} \Delta_2)  G +  (\lambda_{21} GL_1+\lambda_{22} GL_2)
  \end{eqnarray}
where
 \begin{equation}\label{Eq:GX}
 G=2\sum_{n=0}^{N_d} \frac{\pi T}{\omega_n};\;\;\;
 X=\sum_{n=0} \frac{\pi T}{\omega_n^3}
 \end{equation}
 \begin{equation}\label{}
 GL_j=X\left(\frac{v_{Fj}^2}{4} {\bm D}^2 \Delta_j- |\Delta_j|^2\Delta_j\right) \\
 \end{equation}
Expressing $GL_i$ from the equations above we obtain
\begin{eqnarray} \label{Eq:MicroscopicTCGD}
 n_1 GL_1=n_1 \left(\frac{\lambda_{22}}{{\rm Det} \hat\Lambda}-G\right) \Delta_1-
 \frac{\lambda_{J}n_1n_2}{{\rm Det} \hat\Lambda} \Delta_2 \\
 n_2 GL_2=n_2\left(\frac{\lambda_{11}}{{\rm Det} \hat\Lambda}-G\right) \Delta_2-
 \frac{\lambda_{J}n_1n_2}{{\rm Det} \hat\Lambda} \Delta_1
\end{eqnarray}
The system of two coupled GL Eqs.(\ref{Eq:MicroscopicTCGD}) can be obtained minimizing the free energy 
provided the coefficients in Eq.(\ref{Gibbs}) are given by 
\begin{align}\label{Eq:GLexpansion}
 & a_i=  \rho n_i(\tilde{\lambda}_{ii}+\ln T -G_c) \\ \nonumber
 & \gamma = \rho n_1n_2 \lambda_{J}/{\rm Det}\hat \Lambda \\ \nonumber
 & b_i=  \rho n_{i} X/T^2 \\ \nonumber
 & K_i=   v_{Fi}^2b_i/4
 \end{align}
 where $\lambda_J=\lambda_{21}/n_1=\lambda_{12}/n_2$. Note that the expression for $K_i$ in 
 Ref. \cite{Silaev.Babaev:12}
 an extra  coefficient $\rho$. 
  The temperature is normalized to the $T_c$. Here
$X=7\zeta(3)/(8\pi^2) \approx 0.11 $, ${\bar{\lambda}}_{ij}={\lambda}_{ij}^{-1}$
and $G_c=G(T_c)$ is determined by the minimal positive eigenvalue of the inverse coupling matrix $\hat\lambda^{-1}$ :
 $$
 G_c=\frac{{\rm Tr} \lambda-\sqrt{{\rm Tr} \lambda^2-4{\rm Det} \lambda}}{2 {\rm Det}
 \lambda}.
 $$
 We have used the expression $G(T)=G(T_c)-\ln
 T$. Near the critical temperature $\ln T \approx -\tau$ and we obtain
 \begin{eqnarray} \label{Eq:aCoeffitient}
 a_i &=& \alpha_i(T-T_i) \\
 \alpha_i &=& n_i\lambda_{J} \\ 
 T_i &=& (1+G_c-\tilde{\lambda}_{ii}) .
 \end{eqnarray}

In the  above procedure of GL expansion leading to the system (\ref{Eq:MicroscopicTCGD}) 
we assumed both the eigenvalues of the coupling matrix $\hat \lambda$ are positive. 
 

\subsection{Temperature dependence of coherence lengths.}
\label{Sec:Asymptotic}

Coherence lengths are given by the inverse masses of linear modes.
First we investigate the asymptotic behaviour of the
superconducting gaps formulated in terms of the linear modes of the density fields both
in TCGL and microscopic theories described in the previous section.
To find the linear modes we follow the procedure described in 
the section (\ref{asymp}) using the GL model with expansion coefficients  (\ref{Eq:GLexpansion}). 
 Let us set  $K_1=K_2$ which can be accomplished by rescaling the
 fields $\Delta_{1,2}$. Then the corresponding Hessian matrix 
 (\ref{Eq:Hessian}) can be 
 diagonalized with the $k$-independent rotation
 introducing the normal modes 
 $ {\chi}_\beta=U_{\beta i} (\Delta_i - \Delta_{i0})$
 where $\beta=L,H$ and $i=1,2$.
 The rotation matrix $\hat U$ is characterized by
 the mixing angle\citep{johan2,Silaev.Babaev:11} as follows:
 \begin{equation}\label{Eq:MixingAngle}
 \hat U= \begin{pmatrix}
 \cos\theta_L & \sin\theta_L \\
 -\sin\theta_H & \cos\theta_H \
 \end{pmatrix}
 \end{equation}
Note that in accordance with the results of section (\ref{asymp}) the TCGL theory yields 
identical values of two mixing angles $\theta_L = \theta_H = \Theta$ where $\Theta$ is given by the Eq.(\ref{Eq:MixingAngle0}). 
However, in general, outside the region where GL expansion is accurate, the exact microscopic calculation of asymptotic 
yields deviations $\theta_H\neq \theta_L$. This is discussed in Ref \cite{Silaev.Babaev:12}. 
 

The fields $\chi_{L,H}$ corresponding to the linear combinations of $\Delta_{1,2}$ vary at distinct lengths:
 $\xi_H=1/\mu_H$ and $\xi_L=1/\mu_L$. They constitute coherence lengths of the TCGL theory (\ref{Gibbs}) and characterize
 the asymptotic relaxation of the linear combinations of the
 fields $\Delta_{1,2}$, the linear combinations are represented by the composite fields ${\chi}_{L,H}$. 
   
\begin{figure}[htb!]
\centerline{\includegraphics[width=0.5\linewidth]{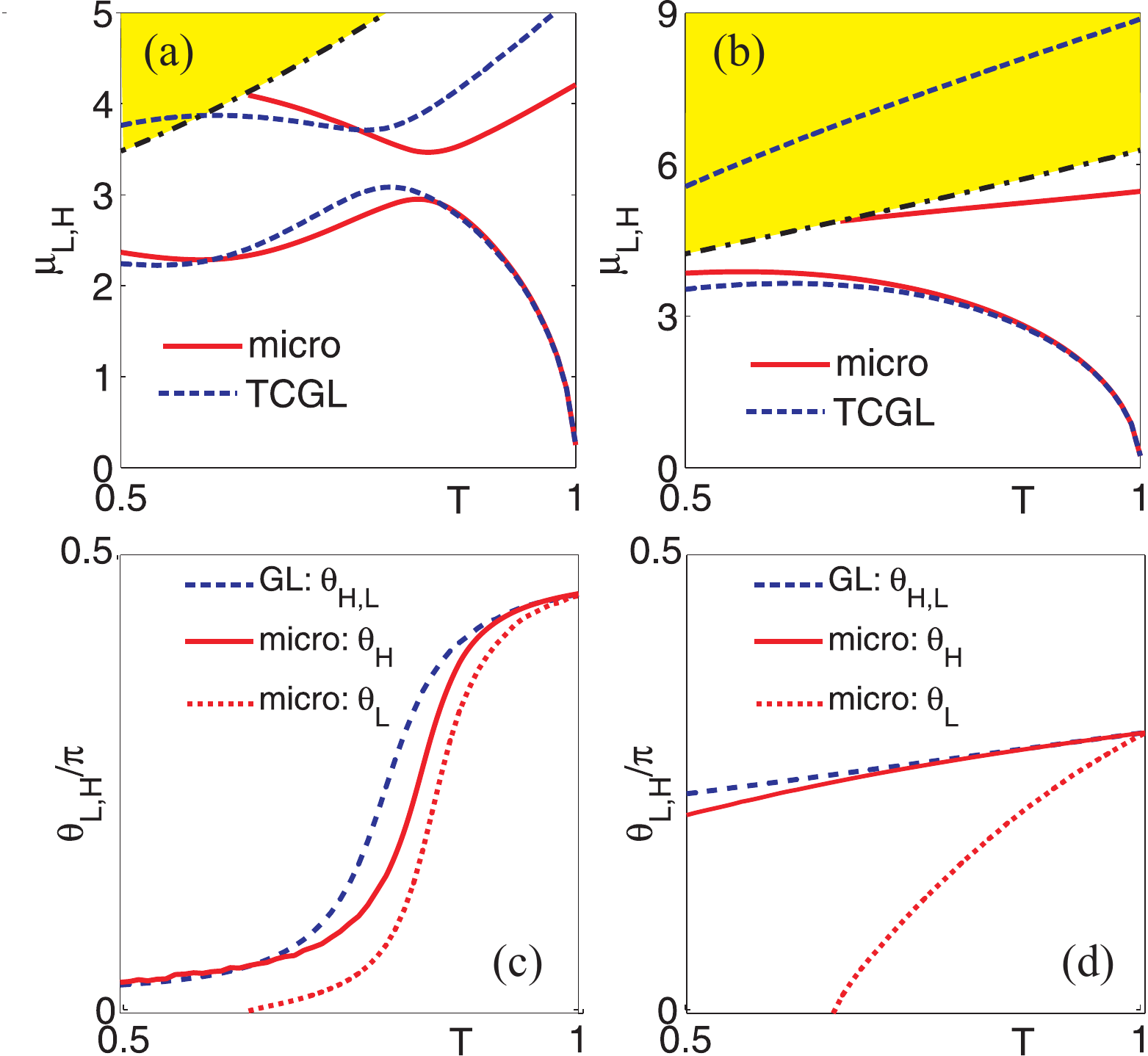}}
\caption{\label{Fig:modes} 
(a) and (b) Comparison of field masses (inverse coherence lengths) given by
full microscopic (solid lines), and microscopically-derived TCGL (dotted) theories. The microscopic parameters are
$\lambda_{11}=0.5$, $\lambda_{22}=0.426$  and
$\lambda_{12}=\lambda_{21}=0.01;\;0.1$ for (a,b)
correspondingly. The yellow shaded region above the dashed-dotted line shows the
continuum of length scales determined by a branch-cut contributions which are specific to the microscopic theory
and  are not captured by  the TCGL description.  
(c,d) Comparison of the mixing angle behaviour
given by the exact microscopic (red lines) and microscopically
derived TCGL theories (blue line). 
 Note that the larger coherence length has a maximum as a function of temperature
deep below $T_c$ near the crossover to the regime when the weak band superconductivity is induced by an interband proximity effect 
 (the corresponding inverse quantity $\mu_L$ has a minimum).
This non-monotonic coherence length behavior is more pronounced at weak interband
coupling and disappears at strong interband coupling \cite{Silaev.Babaev:11}. A multiband system 
with weak interband interaction can easily fall into type-1.5 regime near that crossover temperature.
 Panels (b) and (d) show a
pattern how the TCGL theory  starts to deviate from the
microscopic theory at lower temperature when interband coupling is
increased. Parameters are the same as on
the panels (a,b) correspondingly.}
\end{figure}

 With the help of Eqs.(\ref{Eq:GLexpansion}) for  GL coefficients  obtained from microscopic theory 
 we can study the temperature dependencies of the coherence lengths characterizing the asymptotic relaxation of the gap fields and compare them to the temperature dependence of coherence lengths in full microscopic theory.
Since the system in question breaks only one symmetry, then at critical temperature only one coherence length can diverge
while the second coherence should stay finite. Infinitesimally close to critical temperature $T=T_c-0$ the divergent coherence length 
 has the following standard mean-field behavior $\xi_L=1/\mu_{L}\sim 1/\tau^{1/2}$, where $\tau=1-T/T_c$.
 The contribution of another linear mode in the
 theory sets the scale which is proportional to $\xi_H=1/\mu_{H}$. It remains finite   at $T=T_c$.
 But the amplitude of this mode rapidly vanishes in that region $T=T_c-0$. 
  In Fig.(\ref{Fig:modes})a,b  the temperature dependence of
 masses $\mu_{L,H}$ is plotted comparing the results of the full  microscopic \cite{Silaev.Babaev:11} and
 microscopically derived TCGL theories \cite{Silaev.Babaev:12}.  
 It is shown for the cases of weak and strong interband
coupling in Fig.(\ref{Fig:modes})c,d. We have found that  TCGL
theory describes the lowest characteristic mass $\mu_{L}(T)$ (i.e. the longer coherence length) with
a very good accuracy near $T_c$ [compare the blue and red curves
in Fig. (\ref{Fig:modes})a,b]. Remarkably, when interband coupling
is relatively weak [Fig.(\ref{Fig:modes})c] the ``light" mode is
quite well described by TCGL also at low temperatures down to
$T=0.5 T_c$ around which the weak band crosses over from active to
passive (proximity-induced) superconductivity. Indeed the $\tau$
parameter is large in that case. Nonetheless if the interband coupling is
small one does have a small parameter to implement a GL expansion
for one of the components. Namely one can still expand, e.g. in
the powers of the weak gap $|\Delta_2|/\pi T\ll 1$.
 On the other hand for the
``heavy" mode we naturally obtain some discrepancies even relatively close to $T_c$,
although TCGL theory gives qualitatively correct picture for this
mode when the interband coupling is not too strong. 
More
substantial discrepancies between TCGL and microscopic theories
appear only at lower temperatures or at stronger interband
coupling [Fig.(\ref{Fig:modes})d] where the microscopic response
function has only one pole, while TCGL theory generically has two
poles. Note that these expected deviations concern shorter-range physics
and do not directly affect long-range intervortex forces. In the type-1.5 regime long-range attractive forces
are governed by core-core interaction which range is set by the larger coherence length (lighter mode).

The microscopic two-band GL expansion discussed in this section has straightforward generalization to N-component expansions in N-band $U(1)$ models \cite{garaud2016microscopically}.
For a discussion of microscopic
GL expansion in  more complicated states such as $s+is$ that break multiple symmetries  see\cite{Chubukov2,garaud2016microscopically}. 
   
\section{Systems with generic breakdown of type-1/type-2 dichotomy}

In this section we discuss the simplest situations
of generic  type-1/type-2 dichotomy breakdown.
One example is  superconducting systems that exhibit 
a phase transition from $U(1)$ to $U(1)\times U(1)$ state (or
similar transitions between the states with broken higher symmetries), such 
as the theoretically discussed superconducting states of liquid metallic hydrogen
or deuterium \cite{Nature}, or models involving mixture of protonic and $\Sigma^-$
hyperonic condensates in neutron stars  \cite{Jones21092006},
as well as microscopic certain interface superconductors
\cite{Agterberg.Babaev.ea:14}. Indeed at 
such a transition the magnetic field penetration length remains finite
but there is a divergent coherence length due to the breakdown of additional
symmetry (if the phase transition is continuous). Also the mode associated with the divergent coherence length 
looses its amplitude at the phase transition. Therefore near this 
transition one of the coherence lengths is the largest length scale
of the problem and the system can only be either a type-1 or type-1.5 superconductor.

Similar but more subtle situation takes place at the transition from $s$ to
$s+is$ state \cite{Johan}. The $s+is$ superconductor breaks additional $Z_2$ symmetry
and there is a corresponding diverging coherence length in the problem.
An important generic aspect of the $s+is$ superconducting states 
is that  the density excitations are coupled with the phase
difference excitations in the linear theory \cite{Johan}. 
One of the mixed phase-difference--density mode gives raise to a divergent
coherence length at that phase transition.  Thus such a system  can 
be either   type-1 or type-1.5 near the transition from $s$ to
$s+is$ state. We discuss this example  in more detail in Section VIII.

\section{Structure of the vortex clusters in type-1.5  two-component superconductor.}


In this section, following Ref. \cite{nonpairwise} we consider in more detail the full non-linear problem 
in two-component Ginzburg-Landau models, with
and without Josephson
 coupling $\eta$ which  directly couples the two condensates  (for treatment of other kinds of
interband couplings see \cite{johan2}, for microscopic derivation of the coefficients
see  Sec.\ref{microscopic}). When $\eta=0$ the condensates are
coupled electromagnetically. When there is non-zero interband Josephson coupling, the phase difference is associated with a massive mode with mass
 $\sqrt{\eta (u_1^2+ u_2^2) /u_1 u_2}$.
\begin{align}
\mathcal{F}&=
\frac{1}{2}\sum_{i=1,2}\Biggl[|(\nabla+ ie{\bf A}) \psi_i  |^2
+ (2\alpha_i+\beta_i|\psi_i|^2)|\psi_i|^2\Biggr]
+\frac{1}{2}(\nabla \times {\bf A})^2 -\eta|\psi_1|| \psi_2|\cos(\theta_2-\theta_1)
\label{FreeEnergy}
\end{align}

Since the Ginzburg-Landau model is non-linear, in general intervortex interaction
is non-pairwise. Non-pairwise interaction are important at shorter ranges where 
linearised theory, considered above, does not in general apply.
Below we discuss the importance of  complicated non-pairwise forces between superconducting vortices arising
in certain cases in multicomponent systems \cite{nonpairwise,Garaud.Babaev:15,edstrom2013three}. These non-pairwise forces in certain cases have important consequences for vortex clusters
formation in the type-1.5
regime.


\begin{figure}[!htb]
  \hbox to \linewidth{ \hss
 \includegraphics[width=\linewidth]{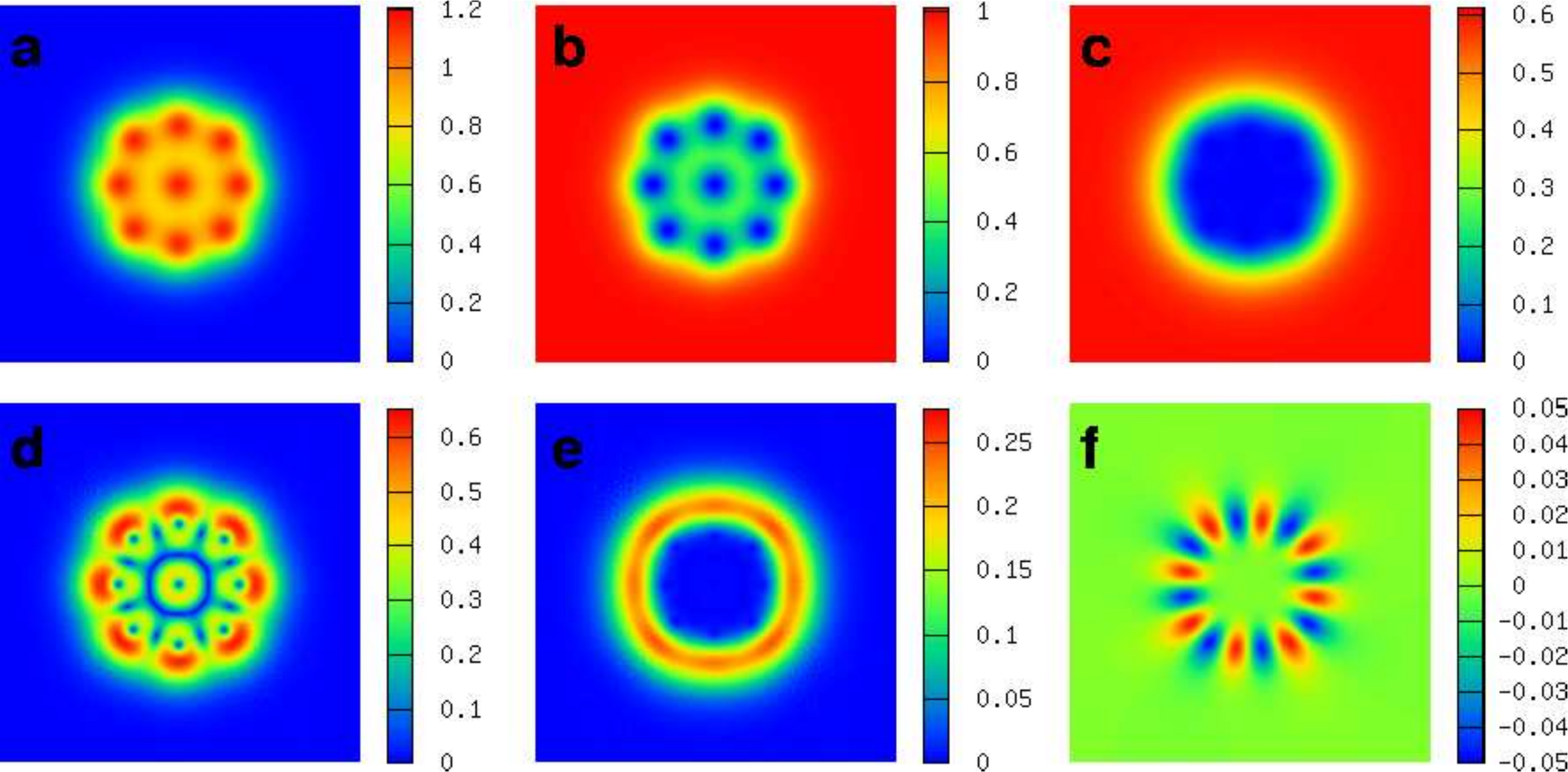}
 \hss}
\caption{
Ground state of $N_v=9$ flux quanta in a  $U(1)\times U(1)$ type-1.5 superconductor 
  (\ie $\eta=0$). The parameters of the potential being
here $(\alpha_1,\beta_1)=(-1.00, 1.00)$ and $(\alpha_2,\beta_2)=(-0.60, 1.00)$, while the electric
charge is $e=1.48$ (in these units the electric charge value parameterises London penetration length).
The displayed physical quantities are $\bf a$ the magnetic flux density,
$\bf b$ (\resp $\bf c$) is the density of the first (\resp second) condensate $|\psi_{1,2}|^2$.
$\bf d$ (\resp $\bf e$) shows the norm of the supercurrent in the first (\resp second) component.
Panel $\bf f$ is $\mathrm{Im}(\psi_1^*\psi_2)\equiv|\psi_1|| \psi_2|\sin(\theta_2-\theta_1)$ being nonzero
when there
appears a difference between the phase of two condensates. 
The solution shows that clearly there is vortex interaction-induced phase-difference gradient
which contributes to non-pairwise intervortex forces.
Parameters are chosen so that the second component
has a type-1 like behavior while the first one tends to from well separated vortices.
The density of the second band is depleted in the vortex cluster and its current is mostly concentrated
on the boundary of the cluster (see Ref.\cite{nonpairwise}).
}
\label{2A-1}
\end{figure}

\begin{figure}[!htb]
  \hbox to \linewidth{ \hss
 \includegraphics[width=\linewidth]{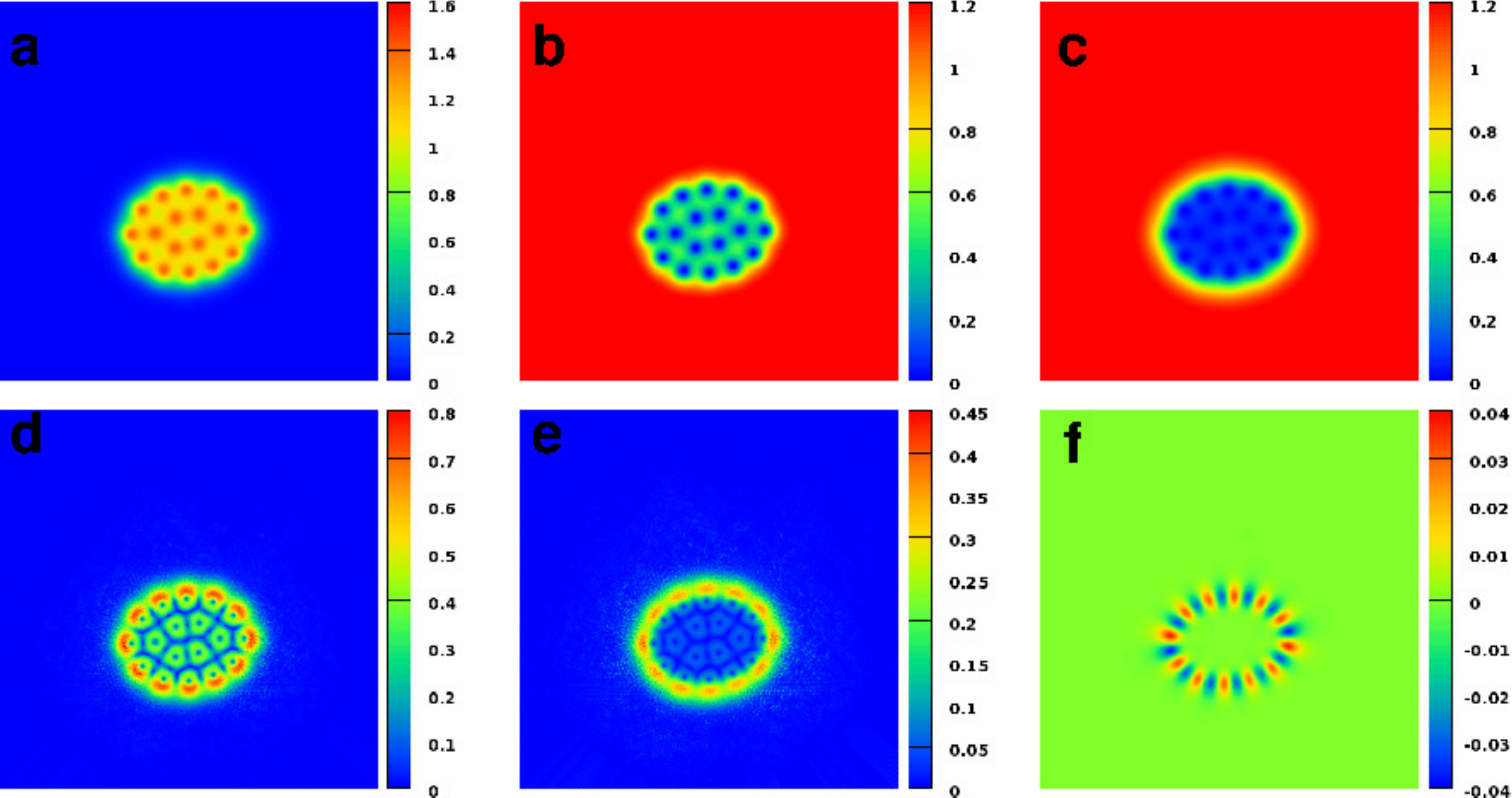}
 \hss}
\caption{
Elongated ground state cluster of $18$ vortices in a superconductor with two active bands. Parameters of the interacting potential
are $(\alpha_1,\beta_1)=(-1.00, 1.00)$, $(\alpha_2,\beta_2)=(-0.0625, 0.25)$ while the interband coupling is $\eta=0.5$.
The electric charge, parameterizing the penetration depth of the magnetic field, is $e=1.30$
so that the well in the nonmonotonic interacting potential is very small. In this case
there is visible admixture of the current of  second component in vortices inside the cluster, though
its current is predominantly concentrated on the boundary of the cluster.
}
\label{fig-new4}
\end{figure}

\Figref{2A-1} and \Figref{fig-new4} show 
numerical solutions for N-vortex bound states in several regimes (for technical details see Appendix of~\cite{nonpairwise}).
The common aspect of the regimes shown on these figures is that the density of
one of the components is depleted in the vortex cluster and
has its current mostly concentrated on the boundary of the vortex cluster (i.e.
has a ``type-1"-like behavior). At the same time the second component
forms a  distinct vortex lattice inside the vortex cluster (i.e. has a ``type-2"-like
behavior).

\begin{figure}[!htb]
  \hbox to \linewidth{ \hss
 \includegraphics[width=\linewidth]{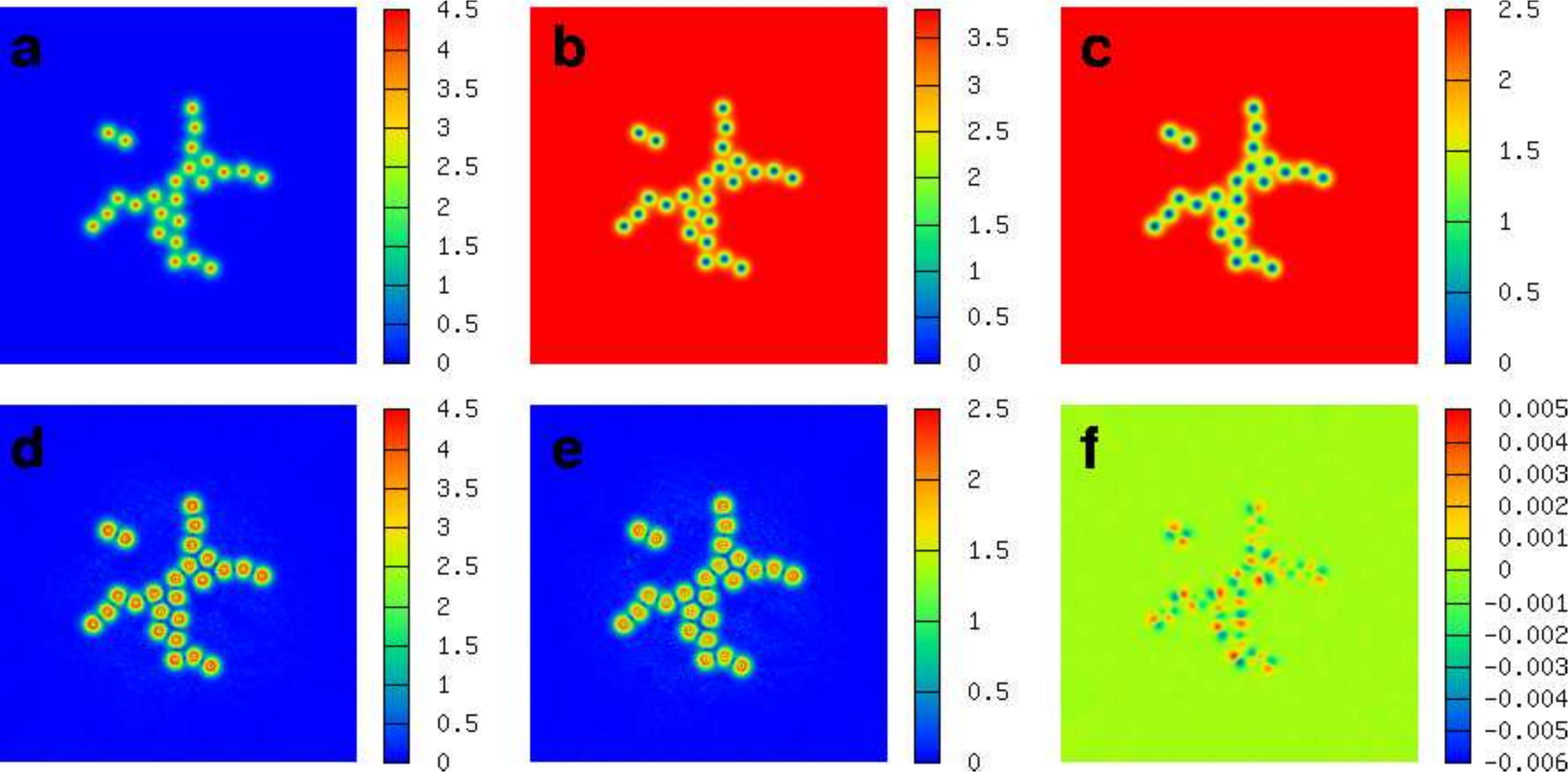}
 \hss}
\caption{A  bound state of an $N_v=25$ vortex configuration in case when
superconductivity in the second band is due to interband proximity
effect
and the  Josephson coupling is relatively strong $\eta=7.0$.
The initial configuration in this simulation was a giant vortex. Other
parameters
are $(\alpha_1,\beta_1)=(-1.00, 1.00)$, $(\alpha_2,\beta_2)=(3.00,
0.50)$,  $e=1.30$.
For the  simulations, like the one shown on Fig. \ref{fig-new4}  the stopping criterion
of energy minimization was when  relative variation of the norm of the gradient of the
GL functional
with respect to all degrees of freedom to be less than $10^{-6}$.
 Here
the situation is
slightly different from that shown on two previous figures. Clearly in the shown above
configuration the ground state  was not reached.
However the interaction potentials a such that
the evolution at the later stages becomes extremely slow.
The number of
energy minimization steps in this case was
order of magnitude larger than what was
required for convergence in the previous regimes.
This signals that in similar systems vortex stripes will be likely to form (relative to more compact vortex clusters) for kinetic
reasons as well as due to thermal fluctuations}
\label{1A1PSJb}
 \end{figure}

Next we report the regime where the  passive  second band (i.e. with positive $\alpha_2$) is coupled
to the first band by  strong Josephson coupling $\eta=7.0$ (shown on \Figref{1A1PSJb}). This coupling
imposes  a strong energy penalty both for disparities of the condensates variations and for the difference
between phases of the condensates. In this regime there are also relatively strong non-pairwise intervortex forces
 favouring chain-like vortex arrangements   compared to compact clusters \cite{nonpairwise}.
We get a relatively flat and complicated energy landscape for multi-vortex configurations and the outcome of the
energy minimization strongly depends on initial configuration.
Simulations whose outcome is compact clusters like \Figref{2A-1} and
\Figref{fig-new4} are clearly ground states, since various initial guesses
lead to similar final configurations. Simulating systems like in \Figref{1A1PSJb}
is less straightforward. Numerical evolution in these systems is extremely slow
because of the complicated energy landscape. The final state strongly depends
on the initial field configuration, indicating the configuration is not the ground state, but a ``quasi-stationary"
bound state with a very slow evolution.
Formation of highly disordered states and vortex chains due to
the short-range nature of the attractive potentials and many-body forces was a common outcome of the
simulation in similar type-1.5 regimes \cite{nonpairwise}, in spite
of the use of a fine numerical grid. 
Another example of systems that  are generally more inclined to posses non-pairwise intervortex forces that promote vortex-stripes are superconducting condensates with  phase separation\cite{Garaud.Babaev:15}.

The \Figref{1A1PSJb} shows the typical non-universal outcome of the energy minimization in
this case. A striking feature here is  formation of vortex stripe-like configuration, in  contrast with the ground state expected from the two-body forces in this system.
Namely the axially symmetric two-body potentials with long range attraction and short-range
repulsion (which we have in this case) do not allow ground states with stripe formation.
In that regime the formation of vortex stripes and small lines is aided by the repulsive non-pairwise intervortex interactions  \cite{nonpairwise}.

Note that even in this regime, the non-linear effects in 
caused by intervortex interaction result in   self-induced gradients of the phase
difference, in spite of the  strong Josephson coupling.

When stray fields are taken into account in thin films, they give repulsive
inter vortex interaction at very long distances, while vortices
can retain attractive interaction at intermediated length scales.
 That gives raise to various hierarchical structures
 such as lattices of vortex clusters or vortex stripes \cite{varney2013hierarchical,2016arXiv160500524M}.
 The study of  dynamics demonstrated that such vortex systems can form vortex glass phase 
 \cite{2016arXiv160500553D}. This is in contrast to type-2 superconductors where vortex glass 
can appear only in the presence of vortex pinning and not in clean samples.

\section{Macroscopic separation in domains of different broken symmetries in type-1.5 superconducting state.}
As discussed above a system with non-monotonic intervortex
interaction potentials allow a state
with macroscopic phase separation
in vortex droplets and Meissner domains.
In type-1.5 superconductors this state
can also represent  a phase separation into domains
of states with different broken symmetries.
In this section we will give two different examples of
how such behavior  arises.

Note that in multicomponent superconductors some symmetries are global (i.e.
associated with the degrees of freedom decoupled from vector potential)
 and some are local  i.e. associated with the degrees of freedom
 coupled to vector potential. As is well known, in the later case the concept of spontaneous symmetry breakdown  is not defined
 the same way as in a system with global symmetry. However below, for brevity
 we will not be making terminological distinctions between local and global symmetries
(detailed discussion of these aspects can be found in  e.g. \cite{frac}).

\subsection{\texorpdfstring{Macroscopic phase separation into $U(1)\times U(1)$ and $U(1)$ domains
in type-1.5 regime.}{Semi-Meissner state as a macroscopic phase separation into U(1)xU(1) and U(1) domains.}}

Consider a  superconductor with broken $U(1) \times U(1)$
symmetry, i.e. a collection of independently conserved
condensates with no intercomponent Josephson coupling. As
discussed above, in the vortex cluster state, in the interior of a vortex droplet,
the superconducting component which has vortices with larger cores
is more depleted. In $U(1) \times U(1)$ system the vortices with
phase windings in different condensates are bound
electromagnetically, resulting in an asymptotically logarithmic
interaction potential with a prefactor proportional to
$|\psi_1|^2|\psi_2|^2/(|\psi_1|^2+|\psi_2|^2)$ \cite{frac}, and
even weaker interaction strength at shorter separations.

Consider now a macroscopically large vortex domain. Even if the
second component there is not completely depleted, its density is
suppressed and as a consequence the binding energy between
vortices with different phase windings ($\Delta \theta_1=2\pi,
\Delta \theta_2=0$) and  ($\Delta \theta_1=0, \Delta
\theta_2=2\pi$) can be arbitrarily small. 
Moreover the vortex ordering energy in the component with more depleted density is small as well. As a result, even tiny thermal fluctuation can drive vortex sublattice melting transition \cite{Nature,sublattice} in a large vortex cluster. In that case the
fractional vortices in weaker the component tear themselves off the
fractional vortices in strong the component and form a disordered
state. Note that the vortex sublattice melting is associated with
the  phase transition from $U(1)\times U(1)$ to $U(1)$ broken symmetries
\cite{Nature,sublattice}. Thus, a macroscopically large  vortex cluster  can realise 
a domain of $U(1)$ phase (associated with the superconducting
state of strong component) immersed in domain of vortexless
$U(1)\times U(1)$ Meissner state. If the magnetic field is
increased, the system will go from the vortex clusters state
(with coexisting $U(1)\times U(1)$ and $U(1)$ domains) to a $U(1)$
vortex state.

\subsection{\texorpdfstring{Macroscopic phase separation in $U(1)$ and $U(1)\times Z_2$ domains
in three band type-1.5 superconductors.}{Semi-Meissner state as a macroscopic phase separation in U(1) and U(1)x Z2 domains
in three band type-1.5 superconductors.}}

In this subsection we discuss an example of vortex clusters in   
three-band superconductors that locally break additional $Z_2$ symmetry forming ``phase-frustrated" states.
Such superconductors also allow the coexistence of domains with different broken
symmetries in the ground state.
The minimal GL free energy functional to model a three-band superconductor  is
\begin{equation}
 F= \frac{1}{2}(\nabla \times \mbf A)^2+ \sum_{i =1,2,3}\frac{1}{2}|\mbf D\psi_i|^2
+\alpha_i|\psi_i|^2+\frac{1}{2}\beta_i|\psi_i|^4
+\sum_{i =1,2,3}\sum_{j>i}\eta_{ij}|\psi_i||\psi_j|\cos(\varphi_{ij}) \,.
\label{freeEnergy}
\end{equation}
Here 
the phase difference between two components are denoted $\varphi_{ij}=\theta_j-\theta_i$.
Microscopic derivations of such models describing $s+is$ superconducting states can be found in \cite{Chubukov2,garaud2016microscopically}. 

 Systems with more than two Josephson-coupled bands can exhibit
{phase frustration} \cite{Ng,StanevTesanovic,Johan,Lin,Chubukov2}. For $\eta_{ij}<0$, a given Josephson interaction energy term is minimal for zero phase difference
 (we then refer to the coupling as ``phase-locking" ), while when  $\eta_{ij}>0$ it is minimal for a phase difference equal to $\pi$
 (we then refer to the coupling as ``phase-antilocking" ). Two-component systems 
 with bilinear Josephson coupling are symmetric with respect to the sign change
$\eta_{ij}\to -\eta_{ij}$ as the phase difference changes by a factor $\pi$, for the system to recover the same interaction. However, in
systems with more than two bands there is generally no such symmetry. For example if a three-band system has $\eta>0$ for all
Josephson interactions, then these terms can not be simultaneously minimized, as this would correspond to all
possible phase differences being equal to $\pi$.

The ground state values of the fields  $|\psi_i|$ and $\varphi_{ij}$ of system \Eqref{freeEnergy} are found by minimizing the potential
energy
\be
\sum_i\Big\{\alpha_i|\psi_i|^2+\frac{1}{2}\beta_i|\psi_i|^4\Big\}
+\sum_{j>i}\eta_{ij}|\psi_i||\psi_j|\cos(\varphi_{ij}).
\label{potential}
\ee
This can however not be done analytically in general, though certain properties
can be derived from qualitative arguments. In terms of the sign of the $\eta$'s, there are four principal situations:

\begin{center}
\begin{tabular}{ c||c|cc }
Case & Sign of $\eta_{12},\eta_{13},\eta_{23}$ & Ground State Phases \\
\hline
1& $- - -$ & $\varphi_1=\varphi_2=\varphi_3$ \\
2& $- - +$ & Frustrated  \\
3& $- + +$ & $\varphi_1=\varphi_2=\varphi_3+\pi$ \\
4& $+ + +$ & Frustrated
 \end{tabular}
\end{center}

The case 2) can result in several ground states. If $|\eta_{23}|\ll |\eta_{12}|,\;|\eta_{13}|$, then the phase differences
are generally $\varphi_{ij}=0$. If on the other hand $|\eta_{12}|,\;|\eta_{13}| \ll |\eta_{23}| $ then $\varphi_{23}=\pi$
and $\varphi_{12}$ is either $0$ or $\pi$. However in certain parameter ranges the resulting state is in fact a ``compromise'' where 
$\varphi_{ij}$ is not an integer multiples of $\pi$.

\begin{figure}[!htb]
 \hbox to \linewidth{ \hss
\includegraphics[width=0.6\linewidth]{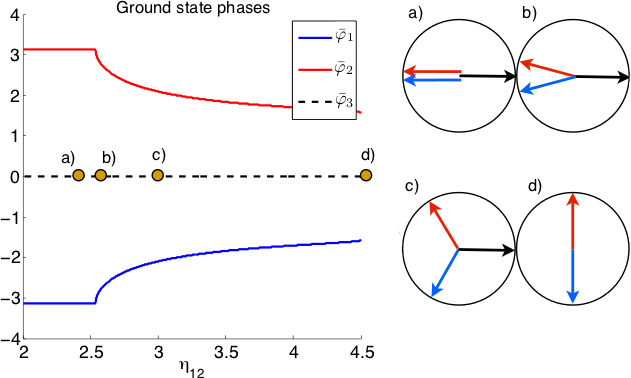}
 \hss}
\caption{
Ground state phases of the three components as function of $\eta_{12}$ (here $\theta_3=0$ fixes the gauge).
The GL parameters are $\alpha_i=1,\;\beta_i=1,\;\eta_{13}=\eta_{23}=3$. For intermediate values of $\eta_{12}$ the ground state
exhibits discrete degeneracy (symmetry is $U(1)\times Z_2$ rather than $U(1)$) since the energy is invariant under the sign
change $\theta_2\to-\theta_2,\; \theta_3\to-\theta_3$. For large $\eta_{12}$ we get $\theta_2-\theta_3=\pi$
implying that $|\psi_3|=0$ and so there is a second transition from $U(1)\times Z_2$ to $U(1)$ and only two bands at the point d).
Here, the phases were computed in a system with only passive bands, though systems with active bands exhibit the same
qualitative properties except for the transition to $U(1)$ and two bands only (\ie active bands have non-zero density in the ground state).
}
\label{case4}
\end{figure}

The case 4) is in fact equivalent to 2) (mapping between these scenarios is trivial). The wide range of resulting groundstates can be seen in \Figref{case4}. As $\eta_{12}$ is scaled,
ground state phases change from $( -\pi,\; \pi,\; 0)$ to the limit where one band is depleted and
the remaining phases are $(-\pi/2,\;\pi/2)$.

An important property of the potential energy \Eqref{potential} is that
if any of the phase differences $\varphi_{ij}$ is not an integer multiple of $\pi$,
then the ground state posses an additional discrete $Z_2$ degeneracy. For example for a system with
$\alpha_i=-1,\;\beta_i=1$ and $\eta_{ij}=1$, two possible ground states exist and are given by
$\varphi_{12}=2\pi/3,\;\varphi_{13}=-2\pi/3$ or $\varphi_{12}=-2\pi/3,\;\varphi_{13}=2\pi/3$.
 Thus in this case, the broken symmetry  is $U(1)\times Z_2$, as opposed to $U(1)$.
As a result, like any other system with $Z_2$  degeneracy, the theory allows an additional set of topological excitations:
domain walls interpolating between the two inequivalent ground states as well as  more complicated topological excitations \cite{Garaud.Carlstrom.ea:11,Garaud.Carlstrom.ea:13,Garaud.Babaev:14}. Generalisations to
frustrated systems with larger number of components was discussed in \cite{Weston.Babaev:13}.

There is a divergent coherence length  at the critical point where
the system undergoes the  $U(1)\times Z_2 \to U(1)$
phase transition (when there is a second order phase transition from an $s+is$ to an $s$ state). The nature of this divergent length-scale is revealed by calculation of the normal modes. Specifically, generating a set of differential equations from the eq.(\ref{freeEnergy}) and linearising these close to the groundstate, gives a mass-matrix whose eigenbasis is also an orthonormal basis of small perturbations to the ground state \cite{Johan}. In systems that break only $U(1)$ symmetry, these modes are segregated with respect to phase and amplitude {so that small perturbations to the phase and amplitude sectors decay independently of each other}. Small perturbations to the amplitude thus has no implications for the phase difference sector, and vice versa. In contrast, in the region where $Z_2$ symmetry is  broken { the modes are generally mixed in this kind of models. In this case a perturbation to the amplitude sector necessarily implies a perturbation to the phase sector} and vice versa. 

The immediate implication of this is that in the region with broken $Z_2$-symmetry, there are five rather than three coherence lengths  that describe amplitude perturbations. If the phase transition
is second order one of  these coherence lengths diverges as we approach the transition point where $Z_2$-symmetry is restored. Thus, vortices in this region produce a perturbation to the amplitude that recovers with a coherence length that is divergent.
Since the magnetic field penetration depth is finite at that transition
the system can be either type-1 or type-1.5
with attractive inter-vortex interaction \cite{Johan}. 
 
In the iron-based superconductors a phase transition from $s$-wave to  $s+is$ state is expected to take place as a function of doping \cite{Chubukov2}. 
Thus, if there is an $s+is$ region at the phase diagram, there should be a  range 
of doping and temperatures in the proximity of the critical point where the type-1.5 superconductivity is generic. 
The general case of $N$-component frustrated superconductors is less studied, however certainly in case of larger number of components there are more possibilities for the appearance of normal modes with low or zero masses leading to type-1.5 regimes \cite{Weston.Babaev:13}. 

\subsection{Non-linear effects and long-range intervortex interaction in $s+is$ superconductors.}
The ground state of a phase frustrated superconductor is in many cases non-trivial, with phase differences being
compromises between the various interaction terms. Inserting vortices in such a system can shift the balance between
different competing couplings, since
vortices can in general have different effects on the different bands. In particular, since the core sizes of vortices
are not generally the same in all bands, vortex matter typically depletes some components more than others and
thus can alter the preferred  values of the phase difference. So the minimal potential energy inside a vortex lattice or
cluster may correspond to a different set of phase differences than in the vortex-free ground state.
 In particular even in $s$-wave systems vortices  can create ``bubbles" of $Z_2$ order parameter around themselves. 
To see this, consider the following argument:
The phase-dependent potential terms in the free energy \Eqref{freeEnergy} are of the form
\be
\eta_{ij} u_iu_j f_i(\mbf r)  f_j(\mbf r)\cos(\varphi_{ij}(\mbf r))\,,
\ee
where $u_i$ are ground state densities and each $f_i(\mbf r)$ represent an Ansatz which models how superfluid densities
are modulated due to vortices. Consider now a  system where N vortices are uniformly distributed in a domain $\Omega$.
The phase dependent part of the free energy is
\be
U_\varphi=\left[\sum_{i>j}\eta_{ij} u_iu_j \right]
  \int_{\Omega} d\mbf r  f_i(\mbf r)  f_j(\mbf r)\cos(\varphi_{ij}(\mbf r)).\label{fullrenorm}
\ee
If $\varphi_{ij}$ is varying slowly in comparison with the inter vortex distance, then it can be considered
constant in a uniform distribution of vortices (as a first approximation). In that case \Eqref{fullrenorm} can be
approximated by
\begin{equation}
U_\varphi\simeq\sum_{ij}\tilde\eta_{ij} u_iu_j\cos(\varphi_{ij})~\text{where}~\tilde{\eta}_{ij}=\eta_{ij}\int_{\Omega} d\mbf r  f_i(\mbf r)  f_j(\mbf r)
\label{renorm}
\end{equation}
If on the other hand $\varphi_{ij}$ varies rapidly, then it is not possible to define $\tilde{\eta}_{ij}$ without a spatial
dependence. Then $\varphi_{ij}$ will depend on $\tilde{\eta}_{ij}(\mbf r)$ which is related to the local modulation functions
$f_if_j$ and vary with a characteristic length scale.

Thus, $\tilde{\eta}$ is the effective inter-band interaction coupling resulting from density modulation.
Since in general, $f_i\not=f_j$ (unless the two bands $i,j$ are identical), one must take into account the
modulation functions $f_i$ when calculating the phase differences. In particular, if the core size in
component $i$ is larger than in component $j$, then  $\int d\mbf r f_if_k <\int d\mbf r f_jf_k$
and therefore the phase differences $\varphi_{ij}$ minimizing \Eqref{renorm}
depend on $f_i$, and consequently on the density of vortices.
Thus  introducing vortices in the system is, in a way, equivalent to a relative effective decrease of some of the Josephson coupling
constants.

\begin{figure}[!htb]
\hbox to \linewidth{ \hss
\includegraphics[width=0.8\linewidth]{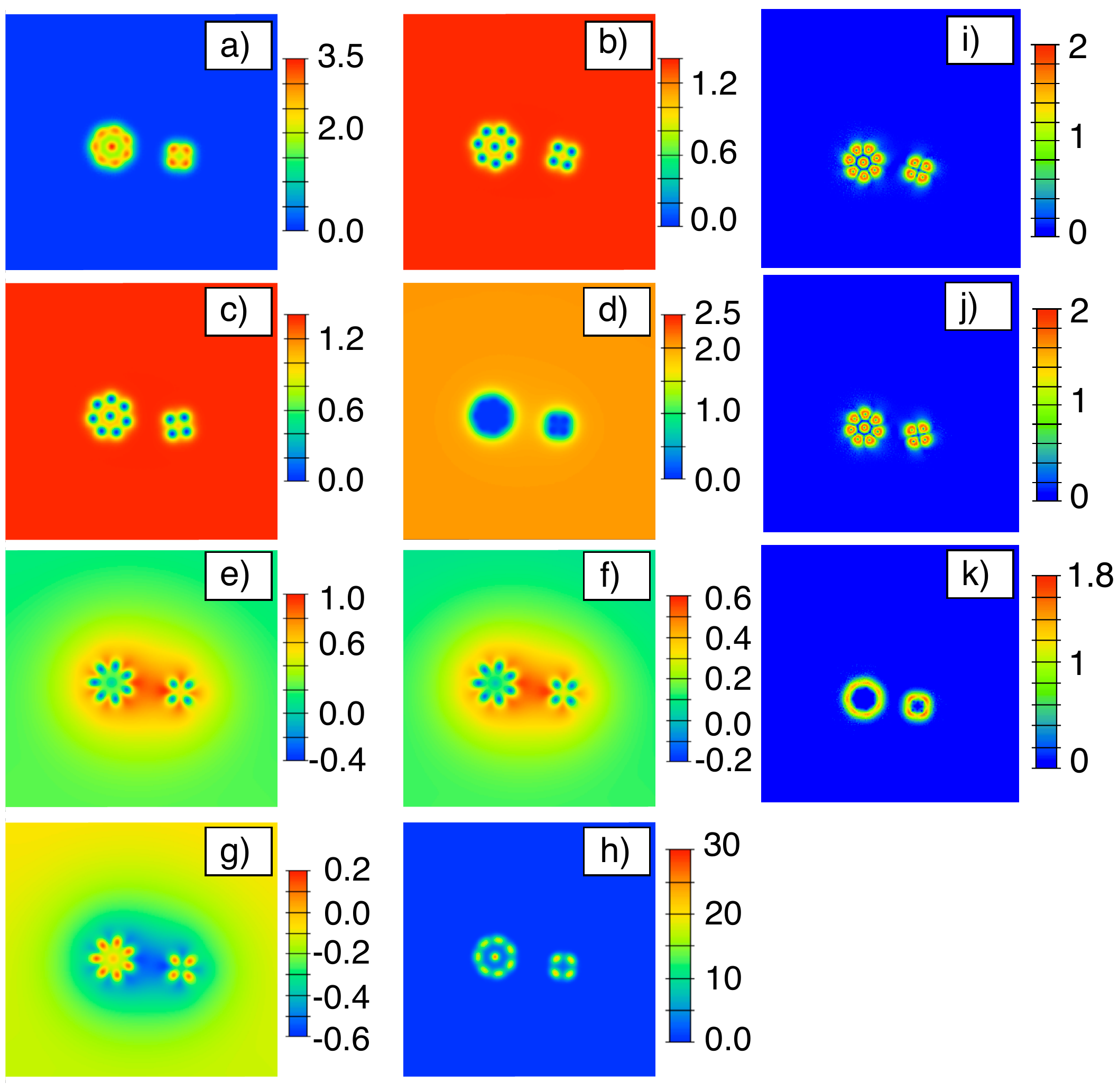}
 \hss}
\caption{
Interacting vortex clusters with internally broken $Z_2$ symmetry in a frustrated three band superconductor. The snapshot
represents a slowly evolving ({quasi-stationary}) state of the weakly interacting well-separated clusters.
In this numerical computation, each of the clusters has with a good accuracy converged to a physical solution
of GL equations, but the snapshot is taken during the slow evolution driven by  the weak long-range
intercluster interaction. The snapshot demonstrates the existence of long-range field variations
associated with the soft mode. The soft mode, as discussed in the text, appears near the $Z_2$
phase transition.
This produces   weak but long-range intervortex forces.
Displayed quantities are: a) Magnetic field, b-d) $|\psi_1|^2,|\psi_2|^2,|\psi_3|^2$, e) $|\psi_1||\psi_2|\sin\varphi_{12}$, f) $|\psi_1||\psi_3|\sin\varphi_{13})$, g) $|\psi_1||\psi_3|\sin\varphi_{23})$.
The GL parameters are $\alpha_1=-3,\;\beta_1=3,\;\alpha_2=-3,\;\beta_2=3, \alpha_3=2,\; \beta_3=0.5,\eta_{12}=2.25,\; \eta_{13}=-3.7$.
The parameter set was chosen so that it lies in the regime where the ground state symmetry of the system without
vortices is $U(1)$, but is close to the $U(1)\times Z_2$ region. Because of the disparity in vortex core size the effective interaction strengths $\tilde{\eta}_{ij}$ are depleted to different extents. As a consequence, 
a vortex cluster produces a bubble of state with broken   $U(1)\times Z_2$ symmetry.
}
\label{ch2}
\end{figure}

\begin{figure}[!htb]
\hbox to \linewidth{ \hss
\includegraphics[width=0.8\linewidth]{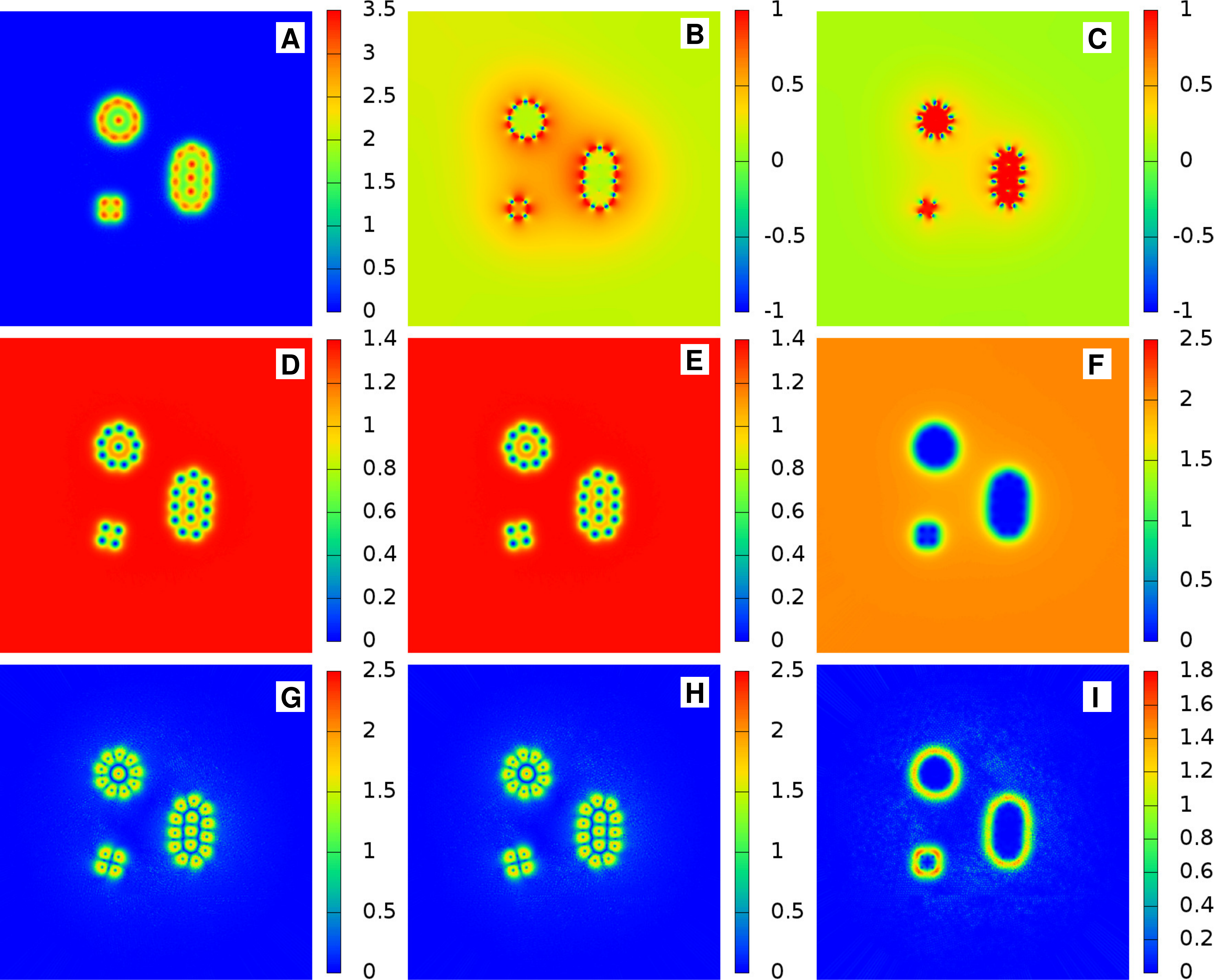}
 \hss}
\caption{
Interacting vortex clusters with broken internal $Z_2$ symmetry in a frustrated three band superconductor.
The panel $\mbf A$  displays the magnetic field ${B}$. Panels $\mbf B$  and $\mbf C$  respectively display $\sin\varphi_{12}$
and $\sin\varphi_{13}$, the third phase difference can obviously be obtained from these two ones. Second line,
shows the densities
of the different condensates $|\psi_1|^2$ ($\mbf D$), $|\psi_2|^2$ ($\mbf E$), $|\psi_3|^2$ ($\mbf F$). The third line displays the
supercurrent densities associated with each condensate $|J_1|$ ($\mbf G$), $|J_2|$ ($\mbf H$), $|J_3|$ ($\mbf I$). The parameter
set here is the same as in \Figref{ch2}.Here the difference compared to the previous picture, is that the sine of the phase differences is represented
`unweighted' by the densities in contrast to \Figref{ch2}, clearly indicating that vortices create an area
with broken $Z_2$ symmetry. Panel $\mbf C$ now makes clear that the inner cluster is in a defined
state $\varphi_{13} \approx \pi/2$ (whose opposite state would have been $-\pi/2$). Panel $\mbf B$ gives a visualization
of the long range interaction between the clusters.
}
\label{c32}
\end{figure}

This can alter the state of the system, as the symmetry of the problem depends on the Josephson interaction terms. In Figs. \ref{ch2} , \ref{c32} we see a type-1.5 system in which the symmetry of the ground state is $U(1)$.
As vortices are inserted into the system, they form clusters and the effective inter band interactions $\tilde{\eta}_{ij}$ are renormalized to a degree
that the symmetry of the domain near vortex clusters changes to $U(1)\times Z_2$.
 Thus the vortex cluster state in such a system represents macroscopic phase separation in domains of broken  $U(1)$  and $U(1)\times Z_2$ symmetries.
 
The vortex structure near the $Z_2$ phase transition which we discussed in this section has important consequences for the phase diagram of the system beyond mean-field approximation, leading to reentrant phase transitions \cite{carlstrom2014spontaneous}. 

In the vicinity of  $Z_2$ phase transition, besides the appearance
of type-1.5 regime, the system has a number of 
other unusual properties such as anomalous vortex viscosity \cite{Silaevvisc}
and distinct anomalous thermoelectric effects \cite{Silaev.Garaud.ea:15,2016PhRvL.116i7002G}.

\section{Fluctuation effects in type-1.5 systems}

In single-component Ginzburg-Landau
models, the order of   superconducting transition
in zero applied magnetic field in three dimensions depends on the ratio of
magnetic field penetration length and coherence length.   Halperin, Lubensky and Ma established that in
extreme type-1 superconductors the gauge field fluctuations make the
superconducting phase transition first order
\cite{HalperinLubenskyMa,COLEMANWEINBERG}.  In the opposite limit of
extreme type-2 systems, Dasgupta and Halperin \cite{DasguptaHalperin}
demonstrated that the superconducting transition is second order in single-component systems and has
the universality class of the inverted-3DXY model.  The  
nature of the superconducting phase transition in this limit is
the proliferation of
vortex-loop  excitations. The inverted-3DXY 
universality class can be demonstrated by duality mapping
\cite{DasguptaHalperin,Peskin1978122,Thomas1978513,ssm}.

The value of the Ginzburg-Landau parameter
$\kappa=\lambda/\xi$ at which the phase transition changes from second
to first order is difficult to establish.   Early numerical 
works suggested that the tricritical point does not coincide
with the Bogomolnyi critical point \cite{bartol}. The
largest Monte Carlo simulations performed at this time
\cite{SudboKappa,SudboKappa2} claim that the tricritical
$\kappa_{\text{tri}}={(0.76\pm0.04)}$ is slightly smaller than
the critical $\kappa_c=1$, which, in our units, separates the type-1 regime
with thermodynamically unstable vortices and the type-2 regime with
thermodynamically stable vortices.  In these works it is claimed that
even in the weakly type-1 regime where the vortex interaction is
purely attractive and vortices are not thermodynamically stable, the
phase transition can be continuous. This raises the question about the 
nature of the phase transition in type-1.5 regime where by contrast 
vortices have long-range attractive interaction but are thermodynamically stable.
The problem was investigated in the effective $j$-current
model \cite{meier} where thermally excited vortices are modeled by
directed loops with long-range attractive, short-range repulsive
interaction similar to the long-range interaction between vortices
in GL model. The results indicate that the zero-field superconducting 
phase transition in type-1.5 materials can be first order \cite{meier}. This is
in contrast to ordinary single-component GL theory which
always has a continuous phase transition in the inverted  3d XY universality class
in the parameter regime where vortices are thermodynamically stable.
For the  $s+is$ type-1.5 systems it was found that fluctuations can modify the 
mean-field phase diagrams quantitatively,  resulting in reentrant phase transitions where
$Z_2$ symmetry is broken by heating \cite{carlstrom2014spontaneous}.

\section{Conclusion}
We briefly outlined the basic concepts and gave a brief account on type-1.5 superconductivity
that takes place in multicomponent systems.
In general a superconducting state state is characterised by multiple coherence lengths
  $\xi_i$, ($i=1,...,M$) arising from multiple broken symmetries or multiple bands. The type-1.5 state is the regime where some of the coherence lengths  are larger
  and some smaller than the magnetic field penetration length:  $\xi_1\leq \xi_2... < \lambda<\xi_M\leq...\leq\xi_N$
  (here we absorbed the usual factor $1/\sqrt{2}$ into the definition of coherence lengths).
  Such a situation should rather generically arise in systems that have phase transitions
  between superconducting states that break different symmetries: at such 
  phase transitions some of the coherence length diverge  while other coherence lengths and magnetic field penetration length stay finite.    Among various unconventional properties that the system acquires in this regime long-range attractive, short-range repulsive intervortex interaction potential.
This allows  a macroscopic phase separation  into domains of Meissner and vortex states in an applied external field. 
This phase separation can also be accompanied  by different broken symmetries in vortex
clusters and Meissner domains.
This regime leads to   unconventional
magnetic, thermal and transport properties.

\section{Acknowledgments}
We thank Julien Garaud for discussions and collaboration on this project.
The work was supported by the Swedish Research Council Grant.
No. 642-2013-7837 and Goran Gustafsson Foundation. J.C. was supported by Wenner-Gren foundation.
The computations were performed on resources provided by the
Swedish National Infrastructure for Computing (SNIC) at National
Supercomputer Center at Link\"oping, Sweden.



\begin{thebibliography}{85}%
\makeatletter
\providecommand \@ifxundefined [1]{%
 \@ifx{#1\undefined}
}%
\providecommand \@ifnum [1]{%
 \ifnum #1\expandafter \@firstoftwo
 \else \expandafter \@secondoftwo
 \fi
}%
\providecommand \@ifx [1]{%
 \ifx #1\expandafter \@firstoftwo
 \else \expandafter \@secondoftwo
 \fi
}%
\providecommand \natexlab [1]{#1}%
\providecommand \enquote  [1]{``#1''}%
\providecommand \bibnamefont  [1]{#1}%
\providecommand \bibfnamefont [1]{#1}%
\providecommand \citenamefont [1]{#1}%
\providecommand \href@noop [0]{\@secondoftwo}%
\providecommand \href [0]{\begingroup \@sanitize@url \@href}%
\providecommand \@href[1]{\@@startlink{#1}\@@href}%
\providecommand \@@href[1]{\endgroup#1\@@endlink}%
\providecommand \@sanitize@url [0]{\catcode `\\12\catcode `\$12\catcode
  `\&12\catcode `\#12\catcode `\^12\catcode `\_12\catcode `\%12\relax}%
\providecommand \@@startlink[1]{}%
\providecommand \@@endlink[0]{}%
\providecommand \url  [0]{\begingroup\@sanitize@url \@url }%
\providecommand \@url [1]{\endgroup\@href {#1}{\urlprefix }}%
\providecommand \urlprefix  [0]{URL }%
\providecommand \Eprint [0]{\href }%
\providecommand \doibase [0]{http://dx.doi.org/}%
\providecommand \selectlanguage [0]{\@gobble}%
\providecommand \bibinfo  [0]{\@secondoftwo}%
\providecommand \bibfield  [0]{\@secondoftwo}%
\providecommand \translation [1]{[#1]}%
\providecommand \BibitemOpen [0]{}%
\providecommand \bibitemStop [0]{}%
\providecommand \bibitemNoStop [0]{.\EOS\space}%
\providecommand \EOS [0]{\spacefactor3000\relax}%
\providecommand \BibitemShut  [1]{\csname bibitem#1\endcsname}%
\let\auto@bib@innerbib\@empty
\bibitem [{\citenamefont {Babaev}\ and\ \citenamefont
  {Speight}(2005)}]{Babaev.Speight:05}%
  \BibitemOpen
  \bibfield  {author} {\bibinfo {author} {\bibfnamefont {E.}~\bibnamefont
  {Babaev}}\ and\ \bibinfo {author} {\bibfnamefont {M.}~\bibnamefont
  {Speight}},\ }\href {\doibase 10.1103/PhysRevB.72.180502} {\bibfield
  {journal} {\bibinfo  {journal} {Phys. Rev. B}\ }\textbf {\bibinfo {volume}
  {72}},\ \bibinfo {pages} {180502} (\bibinfo {year} {2005})}\BibitemShut
  {NoStop}%
\bibitem [{\citenamefont {Landau}\ and\ \citenamefont
  {Ginzburg}(1950)}]{landau1950k}%
  \BibitemOpen
  \bibfield  {author} {\bibinfo {author} {\bibfnamefont {L.}~\bibnamefont
  {Landau}}\ and\ \bibinfo {author} {\bibfnamefont {V.}~\bibnamefont
  {Ginzburg}},\ }\href@noop {} {\bibfield  {journal} {\bibinfo  {journal} {Zh.
  Eksp. Teor. Fiz}\ }\textbf {\bibinfo {volume} {20}},\ \bibinfo {pages} {546}
  (\bibinfo {year} {1950})}\BibitemShut {NoStop}%
\bibitem [{\citenamefont {De~Gennes}(1999)}]{de1999superconductivity}%
  \BibitemOpen
  \bibfield  {author} {\bibinfo {author} {\bibfnamefont {P.}~\bibnamefont
  {De~Gennes}},\ }\href@noop {} {\emph {\bibinfo {title} {Superconductivity of
  Metals and Alloys (Advanced Book Classics)}}}\ (\bibinfo  {publisher}
  {Addison-Wesley Publ. Company Inc},\ \bibinfo {year} {1999})\BibitemShut
  {NoStop}%
\bibitem [{\citenamefont {Abrikosov}(1957)}]{abrikosov1957magnetic}%
  \BibitemOpen
  \bibfield  {author} {\bibinfo {author} {\bibfnamefont {A.~A.}\ \bibnamefont
  {Abrikosov}},\ }\href@noop {} {\bibfield  {journal} {\bibinfo  {journal}
  {Sov. Phys.-JETP (Engl. Transl.);(United States)}\ }\textbf {\bibinfo
  {volume} {5}} (\bibinfo {year} {1957})}\BibitemShut {NoStop}%
\bibitem [{\citenamefont {Kramer}(1971)}]{kramer}%
  \BibitemOpen
  \bibfield  {author} {\bibinfo {author} {\bibfnamefont {L.}~\bibnamefont
  {Kramer}},\ }\href {\doibase 10.1103/PhysRevB.3.3821} {\bibfield  {journal}
  {\bibinfo  {journal} {Phys. Rev. B}\ }\textbf {\bibinfo {volume} {3}},\
  \bibinfo {pages} {3821} (\bibinfo {year} {1971})}\BibitemShut {NoStop}%
\bibitem [{\citenamefont {Bogomol'nyi}(1976)}]{Bogomol}%
  \BibitemOpen
  \bibfield  {author} {\bibinfo {author} {\bibfnamefont {E.}~\bibnamefont
  {Bogomol'nyi}},\ }\href@noop {} {\bibfield  {journal} {\bibinfo  {journal}
  {Sov. J. Nucl. Phys. (Engl. Transl.); (United States)}\ }\textbf {\bibinfo
  {volume} {24:4}} (\bibinfo {year} {1976})}\BibitemShut {NoStop}%
\bibitem [{\citenamefont {Carlstr\"om}\ \emph
  {et~al.}(2011{\natexlab{a}})\citenamefont {Carlstr\"om}, \citenamefont
  {Babaev},\ and\ \citenamefont {Speight}}]{johan2}%
  \BibitemOpen
  \bibfield  {author} {\bibinfo {author} {\bibfnamefont {J.}~\bibnamefont
  {Carlstr\"om}}, \bibinfo {author} {\bibfnamefont {E.}~\bibnamefont {Babaev}},
  \ and\ \bibinfo {author} {\bibfnamefont {M.}~\bibnamefont {Speight}},\ }\href
  {\doibase 10.1103/PhysRevB.83.174509} {\bibfield  {journal} {\bibinfo
  {journal} {Phys. Rev. B}\ }\textbf {\bibinfo {volume} {83}},\ \bibinfo
  {pages} {174509} (\bibinfo {year} {2011}{\natexlab{a}})}\BibitemShut
  {NoStop}%
\bibitem [{\citenamefont {Stanev}\ and\ \citenamefont {Te\ifmmode
  \check{s}\else \v{s}\fi{}anovi\ifmmode~\acute{c}\else
  \'{c}\fi{}}(2010)}]{StanevTesanovic}%
  \BibitemOpen
  \bibfield  {author} {\bibinfo {author} {\bibfnamefont {V.}~\bibnamefont
  {Stanev}}\ and\ \bibinfo {author} {\bibfnamefont {Z.}~\bibnamefont
  {Te\ifmmode \check{s}\else \v{s}\fi{}anovi\ifmmode~\acute{c}\else
  \'{c}\fi{}}},\ }\href {\doibase 10.1103/PhysRevB.81.134522} {\bibfield
  {journal} {\bibinfo  {journal} {Phys. Rev. B}\ }\textbf {\bibinfo {volume}
  {81}},\ \bibinfo {pages} {134522} (\bibinfo {year} {2010})}\BibitemShut
  {NoStop}%
\bibitem [{\citenamefont {Maiti}\ and\ \citenamefont
  {Chubukov}(2013)}]{Chubukov2}%
  \BibitemOpen
  \bibfield  {author} {\bibinfo {author} {\bibfnamefont {S.}~\bibnamefont
  {Maiti}}\ and\ \bibinfo {author} {\bibfnamefont {A.~V.}\ \bibnamefont
  {Chubukov}},\ }\href {\doibase 10.1103/PhysRevB.87.144511} {\bibfield
  {journal} {\bibinfo  {journal} {Phys. Rev. B}\ }\textbf {\bibinfo {volume}
  {87}},\ \bibinfo {pages} {144511} (\bibinfo {year} {2013})}\BibitemShut
  {NoStop}%
\bibitem [{\citenamefont {Carlstr\"om}\ \emph
  {et~al.}(2011{\natexlab{b}})\citenamefont {Carlstr\"om}, \citenamefont
  {Garaud},\ and\ \citenamefont {Babaev}}]{Johan}%
  \BibitemOpen
  \bibfield  {author} {\bibinfo {author} {\bibfnamefont {J.}~\bibnamefont
  {Carlstr\"om}}, \bibinfo {author} {\bibfnamefont {J.}~\bibnamefont {Garaud}},
  \ and\ \bibinfo {author} {\bibfnamefont {E.}~\bibnamefont {Babaev}},\ }\href
  {\doibase 10.1103/PhysRevB.84.134518} {\bibfield  {journal} {\bibinfo
  {journal} {Phys. Rev. B}\ }\textbf {\bibinfo {volume} {84}},\ \bibinfo
  {pages} {134518} (\bibinfo {year} {2011}{\natexlab{b}})}\BibitemShut
  {NoStop}%
\bibitem [{\citenamefont {Babaev}\ \emph {et~al.}(2004)\citenamefont {Babaev},
  \citenamefont {Sudb{\o}},\ and\ \citenamefont {Ashcroft}}]{Nature}%
  \BibitemOpen
  \bibfield  {author} {\bibinfo {author} {\bibfnamefont {E.}~\bibnamefont
  {Babaev}}, \bibinfo {author} {\bibfnamefont {A.}~\bibnamefont {Sudb{\o}}}, \
  and\ \bibinfo {author} {\bibfnamefont {N.}~\bibnamefont {Ashcroft}},\
  }\href@noop {} {\bibfield  {journal} {\bibinfo  {journal} {Nature}\ }\textbf
  {\bibinfo {volume} {431}},\ \bibinfo {pages} {666} (\bibinfo {year}
  {2004})}\BibitemShut {NoStop}%
\bibitem [{\citenamefont {Herland}\ \emph {et~al.}(2010)\citenamefont
  {Herland}, \citenamefont {Babaev},\ and\ \citenamefont
  {Sudb{\o}}}]{herland2010phase}%
  \BibitemOpen
  \bibfield  {author} {\bibinfo {author} {\bibfnamefont {E.~V.}\ \bibnamefont
  {Herland}}, \bibinfo {author} {\bibfnamefont {E.}~\bibnamefont {Babaev}}, \
  and\ \bibinfo {author} {\bibfnamefont {A.}~\bibnamefont {Sudb{\o}}},\
  }\href@noop {} {\bibfield  {journal} {\bibinfo  {journal} {Physical Review
  B}\ }\textbf {\bibinfo {volume} {82}},\ \bibinfo {pages} {134511} (\bibinfo
  {year} {2010})}\BibitemShut {NoStop}%
\bibitem [{\citenamefont {Jones}(2006)}]{Jones21092006}%
  \BibitemOpen
  \bibfield  {author} {\bibinfo {author} {\bibfnamefont {P.~B.}\ \bibnamefont
  {Jones}},\ }\href {\doibase 10.1111/j.1365-2966.2006.10754.x} {\bibfield
  {journal} {\bibinfo  {journal} {Monthly Notices of the Royal Astronomical
  Society}\ }\textbf {\bibinfo {volume} {371}},\ \bibinfo {pages} {1327}
  (\bibinfo {year} {2006})},\ \Eprint
  {http://arxiv.org/abs/http://mnras.oxfordjournals.org/content/371/3/1327.full.pdf+html}
  {http://mnras.oxfordjournals.org/content/371/3/1327.full.pdf+html}
  \BibitemShut {NoStop}%
\bibitem [{\citenamefont {Babaev}(2009)}]{babaev2009unconventional}%
  \BibitemOpen
  \bibfield  {author} {\bibinfo {author} {\bibfnamefont {E.}~\bibnamefont
  {Babaev}},\ }\href@noop {} {\bibfield  {journal} {\bibinfo  {journal}
  {Physical review letters}\ }\textbf {\bibinfo {volume} {103}},\ \bibinfo
  {pages} {231101} (\bibinfo {year} {2009})}\BibitemShut {NoStop}%
\bibitem [{\citenamefont {Suhl}\ \emph {et~al.}(1959)\citenamefont {Suhl},
  \citenamefont {Matthias},\ and\ \citenamefont {Walker}}]{suhl}%
  \BibitemOpen
  \bibfield  {author} {\bibinfo {author} {\bibfnamefont {H.}~\bibnamefont
  {Suhl}}, \bibinfo {author} {\bibfnamefont {B.~T.}\ \bibnamefont {Matthias}},
  \ and\ \bibinfo {author} {\bibfnamefont {L.~R.}\ \bibnamefont {Walker}},\
  }\href {\doibase 10.1103/PhysRevLett.3.552} {\bibfield  {journal} {\bibinfo
  {journal} {Phys. Rev. Lett.}\ }\textbf {\bibinfo {volume} {3}},\ \bibinfo
  {pages} {552} (\bibinfo {year} {1959})}\BibitemShut {NoStop}%
\bibitem [{\citenamefont {Leggett}(1966)}]{LeggettMode}%
  \BibitemOpen
  \bibfield  {author} {\bibinfo {author} {\bibfnamefont {A.~J.}\ \bibnamefont
  {Leggett}},\ }\href {\doibase 10.1143/PTP.36.901} {\bibfield  {journal}
  {\bibinfo  {journal} {Progress of Theoretical Physics}\ }\textbf {\bibinfo
  {volume} {36}},\ \bibinfo {pages} {901} (\bibinfo {year} {1966})}\BibitemShut
  {NoStop}%
\bibitem [{\citenamefont {Tilley}(1964)}]{tilley}%
  \BibitemOpen
  \bibfield  {author} {\bibinfo {author} {\bibfnamefont {D.}~\bibnamefont
  {Tilley}},\ }\href@noop {} {\bibfield  {journal} {\bibinfo  {journal}
  {Proceedings of the Physical Society}\ }\textbf {\bibinfo {volume} {84}},\
  \bibinfo {pages} {573} (\bibinfo {year} {1964})}\BibitemShut {NoStop}%
\bibitem [{\citenamefont {Silaev}\ and\ \citenamefont
  {Babaev}(2012)}]{Silaev.Babaev:12}%
  \BibitemOpen
  \bibfield  {author} {\bibinfo {author} {\bibfnamefont {M.}~\bibnamefont
  {Silaev}}\ and\ \bibinfo {author} {\bibfnamefont {E.}~\bibnamefont
  {Babaev}},\ }\href {\doibase 10.1103/PhysRevB.85.134514} {\bibfield
  {journal} {\bibinfo  {journal} {Phys. Rev. B}\ }\textbf {\bibinfo {volume}
  {85}},\ \bibinfo {pages} {134514} (\bibinfo {year} {2012})}\BibitemShut
  {NoStop}%
\bibitem [{\citenamefont {Garaud}\ \emph {et~al.}(2012)\citenamefont {Garaud},
  \citenamefont {Agterberg},\ and\ \citenamefont
  {Babaev}}]{Garaud.Agterberg.ea:12}%
  \BibitemOpen
  \bibfield  {author} {\bibinfo {author} {\bibfnamefont {J.}~\bibnamefont
  {Garaud}}, \bibinfo {author} {\bibfnamefont {D.~F.}\ \bibnamefont
  {Agterberg}}, \ and\ \bibinfo {author} {\bibfnamefont {E.}~\bibnamefont
  {Babaev}},\ }\href {\doibase 10.1103/PhysRevB.86.060513} {\bibfield
  {journal} {\bibinfo  {journal} {Phys. Rev. B}\ }\textbf {\bibinfo {volume}
  {86}},\ \bibinfo {pages} {060513} (\bibinfo {year} {2012})}\BibitemShut
  {NoStop}%
\bibitem [{\citenamefont {Frank}\ and\ \citenamefont {Lemm}(2016)}]{Frank2016}%
  \BibitemOpen
  \bibfield  {author} {\bibinfo {author} {\bibfnamefont {R.~L.}\ \bibnamefont
  {Frank}}\ and\ \bibinfo {author} {\bibfnamefont {M.}~\bibnamefont {Lemm}},\
  }\href {\doibase 10.1007/s00023-016-0473-x} {\bibfield  {journal} {\bibinfo
  {journal} {Annales Henri Poincar{\'e}}\ ,\ \bibinfo {pages} {1}} (\bibinfo
  {year} {2016})}\BibitemShut {NoStop}%
\bibitem [{\citenamefont {Babaev}\ \emph {et~al.}(2010)\citenamefont {Babaev},
  \citenamefont {Carlstr\"om},\ and\ \citenamefont {Speight}}]{johan1}%
  \BibitemOpen
  \bibfield  {author} {\bibinfo {author} {\bibfnamefont {E.}~\bibnamefont
  {Babaev}}, \bibinfo {author} {\bibfnamefont {J.}~\bibnamefont {Carlstr\"om}},
  \ and\ \bibinfo {author} {\bibfnamefont {M.}~\bibnamefont {Speight}},\ }\href
  {\doibase 10.1103/PhysRevLett.105.067003} {\bibfield  {journal} {\bibinfo
  {journal} {Phys. Rev. Lett.}\ }\textbf {\bibinfo {volume} {105}},\ \bibinfo
  {pages} {067003} (\bibinfo {year} {2010})}\BibitemShut {NoStop}%
\bibitem [{\citenamefont {Silaev}\ and\ \citenamefont
  {Babaev}(2011)}]{Silaev.Babaev:11}%
  \BibitemOpen
  \bibfield  {author} {\bibinfo {author} {\bibfnamefont {M.}~\bibnamefont
  {Silaev}}\ and\ \bibinfo {author} {\bibfnamefont {E.}~\bibnamefont
  {Babaev}},\ }\href {\doibase 10.1103/PhysRevB.84.094515} {\bibfield
  {journal} {\bibinfo  {journal} {Phys. Rev. B}\ }\textbf {\bibinfo {volume}
  {84}},\ \bibinfo {pages} {094515} (\bibinfo {year} {2011})}\BibitemShut
  {NoStop}%
\bibitem [{\citenamefont {{Carlstr{\"o}m}}\ \emph {et~al.}(2011)\citenamefont
  {{Carlstr{\"o}m}}, \citenamefont {{Garaud}},\ and\ \citenamefont
  {{Babaev}}}]{nonpairwise}%
  \BibitemOpen
  \bibfield  {author} {\bibinfo {author} {\bibfnamefont {J.}~\bibnamefont
  {{Carlstr{\"o}m}}}, \bibinfo {author} {\bibfnamefont {J.}~\bibnamefont
  {{Garaud}}}, \ and\ \bibinfo {author} {\bibfnamefont {E.}~\bibnamefont
  {{Babaev}}},\ }\href {\doibase 10.1103/PhysRevB.84.134515} {\bibfield
  {journal} {\bibinfo  {journal} {\prb}\ }\textbf {\bibinfo {volume} {84}},\
  \bibinfo {eid} {134515} (\bibinfo {year} {2011})}\BibitemShut {NoStop}%
\bibitem [{\citenamefont {{Moshchalkov}}\ \emph {et~al.}(2009)\citenamefont
  {{Moshchalkov}}, \citenamefont {{Menghini}}, \citenamefont {{Nishio}},
  \citenamefont {{Chen}}, \citenamefont {{Silhanek}}, \citenamefont {{Dao}},
  \citenamefont {{Chibotaru}}, \citenamefont {{Zhigadlo}},\ and\ \citenamefont
  {{Karpinski}}}]{moshchalkov}%
  \BibitemOpen
  \bibfield  {author} {\bibinfo {author} {\bibfnamefont {V.}~\bibnamefont
  {{Moshchalkov}}}, \bibinfo {author} {\bibfnamefont {M.}~\bibnamefont
  {{Menghini}}}, \bibinfo {author} {\bibfnamefont {T.}~\bibnamefont
  {{Nishio}}}, \bibinfo {author} {\bibfnamefont {Q.~H.}\ \bibnamefont
  {{Chen}}}, \bibinfo {author} {\bibfnamefont {A.~V.}\ \bibnamefont
  {{Silhanek}}}, \bibinfo {author} {\bibfnamefont {V.~H.}\ \bibnamefont
  {{Dao}}}, \bibinfo {author} {\bibfnamefont {L.~F.}\ \bibnamefont
  {{Chibotaru}}}, \bibinfo {author} {\bibfnamefont {N.~D.}\ \bibnamefont
  {{Zhigadlo}}}, \ and\ \bibinfo {author} {\bibfnamefont {J.}~\bibnamefont
  {{Karpinski}}},\ }\href {\doibase 10.1103/PhysRevLett.102.117001} {\bibfield
  {journal} {\bibinfo  {journal} {Phys. Rev. Lett.}\ }\textbf {\bibinfo
  {volume} {102}},\ \bibinfo {eid} {117001} (\bibinfo {year}
  {2009})}\BibitemShut {NoStop}%
\bibitem [{\citenamefont {{Nishio}}\ \emph {et~al.}(2010)\citenamefont
  {{Nishio}}, \citenamefont {{Dao}}, \citenamefont {{Chen}}, \citenamefont
  {{Chibotaru}}, \citenamefont {{Kadowaki}},\ and\ \citenamefont
  {{Moshchalkov}}}]{moshchalkov2}%
  \BibitemOpen
  \bibfield  {author} {\bibinfo {author} {\bibfnamefont {T.}~\bibnamefont
  {{Nishio}}}, \bibinfo {author} {\bibfnamefont {V.~H.}\ \bibnamefont {{Dao}}},
  \bibinfo {author} {\bibfnamefont {Q.}~\bibnamefont {{Chen}}}, \bibinfo
  {author} {\bibfnamefont {L.~F.}\ \bibnamefont {{Chibotaru}}}, \bibinfo
  {author} {\bibfnamefont {K.}~\bibnamefont {{Kadowaki}}}, \ and\ \bibinfo
  {author} {\bibfnamefont {V.~V.}\ \bibnamefont {{Moshchalkov}}},\ }\href
  {\doibase 10.1103/PhysRevB.81.020506} {\bibfield  {journal} {\bibinfo
  {journal} {\prb}\ }\textbf {\bibinfo {volume} {81}},\ \bibinfo {eid} {020506}
  (\bibinfo {year} {2010})}\BibitemShut {NoStop}%
\bibitem [{\citenamefont {Hicks}\ \emph {et~al.}(2010)\citenamefont {Hicks},
  \citenamefont {Kirtley}, \citenamefont {Lippman}, \citenamefont {Koshnick},
  \citenamefont {Huber}, \citenamefont {Maeno}, \citenamefont {Yuhasz},
  \citenamefont {Maple},\ and\ \citenamefont {Moler}}]{moler2}%
  \BibitemOpen
  \bibfield  {author} {\bibinfo {author} {\bibfnamefont {C.~W.}\ \bibnamefont
  {Hicks}}, \bibinfo {author} {\bibfnamefont {J.~R.}\ \bibnamefont {Kirtley}},
  \bibinfo {author} {\bibfnamefont {T.~M.}\ \bibnamefont {Lippman}}, \bibinfo
  {author} {\bibfnamefont {N.~C.}\ \bibnamefont {Koshnick}}, \bibinfo {author}
  {\bibfnamefont {M.~E.}\ \bibnamefont {Huber}}, \bibinfo {author}
  {\bibfnamefont {Y.}~\bibnamefont {Maeno}}, \bibinfo {author} {\bibfnamefont
  {W.~M.}\ \bibnamefont {Yuhasz}}, \bibinfo {author} {\bibfnamefont {M.~B.}\
  \bibnamefont {Maple}}, \ and\ \bibinfo {author} {\bibfnamefont {K.~A.}\
  \bibnamefont {Moler}},\ }\href {\doibase 10.1103/PhysRevB.81.214501}
  {\bibfield  {journal} {\bibinfo  {journal} {Phys. Rev. B}\ }\textbf {\bibinfo
  {volume} {81}},\ \bibinfo {pages} {214501} (\bibinfo {year}
  {2010})}\BibitemShut {NoStop}%
\bibitem [{\citenamefont {Ray}\ \emph {et~al.}(2014)\citenamefont {Ray},
  \citenamefont {Gibbs}, \citenamefont {Bending}, \citenamefont {Curran},
  \citenamefont {Babaev}, \citenamefont {Baines}, \citenamefont {Mackenzie},\
  and\ \citenamefont {Lee}}]{Ray.Gibbs.ea:14}%
  \BibitemOpen
  \bibfield  {author} {\bibinfo {author} {\bibfnamefont {S.~J.}\ \bibnamefont
  {Ray}}, \bibinfo {author} {\bibfnamefont {A.~S.}\ \bibnamefont {Gibbs}},
  \bibinfo {author} {\bibfnamefont {S.~J.}\ \bibnamefont {Bending}}, \bibinfo
  {author} {\bibfnamefont {P.~J.}\ \bibnamefont {Curran}}, \bibinfo {author}
  {\bibfnamefont {E.}~\bibnamefont {Babaev}}, \bibinfo {author} {\bibfnamefont
  {C.}~\bibnamefont {Baines}}, \bibinfo {author} {\bibfnamefont {A.~P.}\
  \bibnamefont {Mackenzie}}, \ and\ \bibinfo {author} {\bibfnamefont {S.~L.}\
  \bibnamefont {Lee}},\ }\href {\doibase 10.1103/PhysRevB.89.094504} {\bibfield
   {journal} {\bibinfo  {journal} {Phys. Rev. B}\ }\textbf {\bibinfo {volume}
  {89}},\ \bibinfo {pages} {094504} (\bibinfo {year} {2014})}\BibitemShut
  {NoStop}%
\bibitem [{\citenamefont {Kawasaki}\ \emph {et~al.}(2013)\citenamefont
  {Kawasaki}, \citenamefont {Watanabe}, \citenamefont {Amitsuka}, \citenamefont
  {Kunimori}, \citenamefont {Tanida},\ and\ \citenamefont
  {{\=O}nuki}}]{noncentr}%
  \BibitemOpen
  \bibfield  {author} {\bibinfo {author} {\bibfnamefont {I.}~\bibnamefont
  {Kawasaki}}, \bibinfo {author} {\bibfnamefont {I.}~\bibnamefont {Watanabe}},
  \bibinfo {author} {\bibfnamefont {H.}~\bibnamefont {Amitsuka}}, \bibinfo
  {author} {\bibfnamefont {K.}~\bibnamefont {Kunimori}}, \bibinfo {author}
  {\bibfnamefont {H.}~\bibnamefont {Tanida}}, \ and\ \bibinfo {author}
  {\bibfnamefont {Y.}~\bibnamefont {{\=O}nuki}},\ }\href {\doibase
  10.7566/JPSJ.82.084713} {\bibfield  {journal} {\bibinfo  {journal} {Journal
  of the Physical Society of Japan}\ }\textbf {\bibinfo {volume} {82}},\
  \bibinfo {pages} {084713} (\bibinfo {year} {2013})}\BibitemShut {NoStop}%
\bibitem [{\citenamefont {Fujisawa}\ \emph {et~al.}(2015)\citenamefont
  {Fujisawa}, \citenamefont {Yamaguchi}, \citenamefont {Motoyama},
  \citenamefont {Kawakatsu}, \citenamefont {Sumiyama}, \citenamefont
  {Takeuchi}, \citenamefont {Settai},\ and\ \citenamefont
  {{\AA}nuki}}]{1347-4065-54-4-048001}%
  \BibitemOpen
  \bibfield  {author} {\bibinfo {author} {\bibfnamefont {T.}~\bibnamefont
  {Fujisawa}}, \bibinfo {author} {\bibfnamefont {A.}~\bibnamefont {Yamaguchi}},
  \bibinfo {author} {\bibfnamefont {G.}~\bibnamefont {Motoyama}}, \bibinfo
  {author} {\bibfnamefont {D.}~\bibnamefont {Kawakatsu}}, \bibinfo {author}
  {\bibfnamefont {A.}~\bibnamefont {Sumiyama}}, \bibinfo {author}
  {\bibfnamefont {T.}~\bibnamefont {Takeuchi}}, \bibinfo {author}
  {\bibfnamefont {R.}~\bibnamefont {Settai}}, \ and\ \bibinfo {author}
  {\bibfnamefont {Y.}~\bibnamefont {{\AA}nuki}},\ }\href
  {http://stacks.iop.org/1347-4065/54/i=4/a=048001} {\bibfield  {journal}
  {\bibinfo  {journal} {Japanese Journal of Applied Physics}\ }\textbf
  {\bibinfo {volume} {54}},\ \bibinfo {pages} {048001} (\bibinfo {year}
  {2015})}\BibitemShut {NoStop}%
\bibitem [{\citenamefont {Dumont}\ and\ \citenamefont
  {Mota}(2002)}]{PhysRevB.65.144519}%
  \BibitemOpen
  \bibfield  {author} {\bibinfo {author} {\bibfnamefont {E.}~\bibnamefont
  {Dumont}}\ and\ \bibinfo {author} {\bibfnamefont {A.~C.}\ \bibnamefont
  {Mota}},\ }\href {\doibase 10.1103/PhysRevB.65.144519} {\bibfield  {journal}
  {\bibinfo  {journal} {Phys. Rev. B}\ }\textbf {\bibinfo {volume} {65}},\
  \bibinfo {pages} {144519} (\bibinfo {year} {2002})}\BibitemShut {NoStop}%
\bibitem [{\citenamefont {Garaud}\ \emph {et~al.}(2016)\citenamefont {Garaud},
  \citenamefont {Babaev}, \citenamefont {Bojesen},\ and\ \citenamefont
  {Sudb{\o}}}]{garaud2016lattices}%
  \BibitemOpen
  \bibfield  {author} {\bibinfo {author} {\bibfnamefont {J.}~\bibnamefont
  {Garaud}}, \bibinfo {author} {\bibfnamefont {E.}~\bibnamefont {Babaev}},
  \bibinfo {author} {\bibfnamefont {T.~A.}\ \bibnamefont {Bojesen}}, \ and\
  \bibinfo {author} {\bibfnamefont {A.}~\bibnamefont {Sudb{\o}}},\ }\href@noop
  {} {\bibfield  {journal} {\bibinfo  {journal} {arXiv preprint
  arXiv:1605.03946}\ } (\bibinfo {year} {2016})}\BibitemShut {NoStop}%
\bibitem [{\citenamefont {Garaud}\ and\ \citenamefont
  {Babaev}(2015{\natexlab{a}})}]{garaud2015properties}%
  \BibitemOpen
  \bibfield  {author} {\bibinfo {author} {\bibfnamefont {J.}~\bibnamefont
  {Garaud}}\ and\ \bibinfo {author} {\bibfnamefont {E.}~\bibnamefont
  {Babaev}},\ }\href@noop {} {\bibfield  {journal} {\bibinfo  {journal}
  {Scientific reports}\ }\textbf {\bibinfo {volume} {5}} (\bibinfo {year}
  {2015}{\natexlab{a}})}\BibitemShut {NoStop}%
\bibitem [{\citenamefont {Agterberg}\ \emph {et~al.}(2014)\citenamefont
  {Agterberg}, \citenamefont {Babaev},\ and\ \citenamefont
  {Garaud}}]{Agterberg.Babaev.ea:14}%
  \BibitemOpen
  \bibfield  {author} {\bibinfo {author} {\bibfnamefont {D.~F.}\ \bibnamefont
  {Agterberg}}, \bibinfo {author} {\bibfnamefont {E.}~\bibnamefont {Babaev}}, \
  and\ \bibinfo {author} {\bibfnamefont {J.}~\bibnamefont {Garaud}},\ }\href
  {\doibase 10.1103/PhysRevB.90.064509} {\bibfield  {journal} {\bibinfo
  {journal} {Phys. Rev. B}\ }\textbf {\bibinfo {volume} {90}},\ \bibinfo
  {pages} {064509} (\bibinfo {year} {2014})}\BibitemShut {NoStop}%
\bibitem [{\citenamefont {Parameswaran}\ \emph {et~al.}(2012)\citenamefont
  {Parameswaran}, \citenamefont {Kivelson}, \citenamefont {Rezayi},
  \citenamefont {Simon}, \citenamefont {Sondhi},\ and\ \citenamefont
  {Spivak}}]{parameswaran2012typology}%
  \BibitemOpen
  \bibfield  {author} {\bibinfo {author} {\bibfnamefont {S.}~\bibnamefont
  {Parameswaran}}, \bibinfo {author} {\bibfnamefont {S.}~\bibnamefont
  {Kivelson}}, \bibinfo {author} {\bibfnamefont {E.}~\bibnamefont {Rezayi}},
  \bibinfo {author} {\bibfnamefont {S.}~\bibnamefont {Simon}}, \bibinfo
  {author} {\bibfnamefont {S.}~\bibnamefont {Sondhi}}, \ and\ \bibinfo {author}
  {\bibfnamefont {B.}~\bibnamefont {Spivak}},\ }\href@noop {} {\bibfield
  {journal} {\bibinfo  {journal} {Physical Review B}\ }\textbf {\bibinfo
  {volume} {85}},\ \bibinfo {pages} {241307} (\bibinfo {year}
  {2012})}\BibitemShut {NoStop}%
\bibitem [{\citenamefont {Alford}\ and\ \citenamefont
  {Good}(2008)}]{alford2008flux}%
  \BibitemOpen
  \bibfield  {author} {\bibinfo {author} {\bibfnamefont {M.~G.}\ \bibnamefont
  {Alford}}\ and\ \bibinfo {author} {\bibfnamefont {G.}~\bibnamefont {Good}},\
  }\href@noop {} {\bibfield  {journal} {\bibinfo  {journal} {Physical Review
  B}\ }\textbf {\bibinfo {volume} {78}},\ \bibinfo {pages} {024510} (\bibinfo
  {year} {2008})}\BibitemShut {NoStop}%
\bibitem [{\citenamefont {{Dao}}\ \emph {et~al.}(2011)\citenamefont {{Dao}},
  \citenamefont {{Chibotaru}}, \citenamefont {{Nishio}},\ and\ \citenamefont
  {{Moshchalkov}}}]{dao}%
  \BibitemOpen
  \bibfield  {author} {\bibinfo {author} {\bibfnamefont {V.~H.}\ \bibnamefont
  {{Dao}}}, \bibinfo {author} {\bibfnamefont {L.~F.}\ \bibnamefont
  {{Chibotaru}}}, \bibinfo {author} {\bibfnamefont {T.}~\bibnamefont
  {{Nishio}}}, \ and\ \bibinfo {author} {\bibfnamefont {V.~V.}\ \bibnamefont
  {{Moshchalkov}}},\ }\href {\doibase 10.1103/PhysRevB.83.020503} {\bibfield
  {journal} {\bibinfo  {journal} {\prb}\ }\textbf {\bibinfo {volume} {83}},\
  \bibinfo {eid} {020503} (\bibinfo {year} {2011})}\BibitemShut {NoStop}%
\bibitem [{\citenamefont {Gutierrez}\ \emph {et~al.}(2012)\citenamefont
  {Gutierrez}, \citenamefont {Raes}, \citenamefont {Silhanek}, \citenamefont
  {Li}, \citenamefont {Zhigadlo}, \citenamefont {Karpinski}, \citenamefont
  {Tempere},\ and\ \citenamefont {Moshchalkov}}]{gutierrez2012scanning}%
  \BibitemOpen
  \bibfield  {author} {\bibinfo {author} {\bibfnamefont {J.}~\bibnamefont
  {Gutierrez}}, \bibinfo {author} {\bibfnamefont {B.}~\bibnamefont {Raes}},
  \bibinfo {author} {\bibfnamefont {A.}~\bibnamefont {Silhanek}}, \bibinfo
  {author} {\bibfnamefont {L.}~\bibnamefont {Li}}, \bibinfo {author}
  {\bibfnamefont {N.}~\bibnamefont {Zhigadlo}}, \bibinfo {author}
  {\bibfnamefont {J.}~\bibnamefont {Karpinski}}, \bibinfo {author}
  {\bibfnamefont {J.}~\bibnamefont {Tempere}}, \ and\ \bibinfo {author}
  {\bibfnamefont {V.}~\bibnamefont {Moshchalkov}},\ }\href@noop {} {\bibfield
  {journal} {\bibinfo  {journal} {Physical Review B}\ }\textbf {\bibinfo
  {volume} {85}},\ \bibinfo {pages} {094511} (\bibinfo {year}
  {2012})}\BibitemShut {NoStop}%
\bibitem [{\citenamefont {Li}\ \emph {et~al.}(2011)\citenamefont {Li},
  \citenamefont {Nishio}, \citenamefont {Xu},\ and\ \citenamefont
  {Moshchalkov}}]{li2011low}%
  \BibitemOpen
  \bibfield  {author} {\bibinfo {author} {\bibfnamefont {L.}~\bibnamefont
  {Li}}, \bibinfo {author} {\bibfnamefont {T.}~\bibnamefont {Nishio}}, \bibinfo
  {author} {\bibfnamefont {Z.}~\bibnamefont {Xu}}, \ and\ \bibinfo {author}
  {\bibfnamefont {V.}~\bibnamefont {Moshchalkov}},\ }\href@noop {} {\bibfield
  {journal} {\bibinfo  {journal} {Physical Review B}\ }\textbf {\bibinfo
  {volume} {83}},\ \bibinfo {pages} {224522} (\bibinfo {year}
  {2011})}\BibitemShut {NoStop}%
\bibitem [{\citenamefont {Varney}\ \emph {et~al.}(2013)\citenamefont {Varney},
  \citenamefont {Sellin}, \citenamefont {Wang}, \citenamefont {Fangohr},\ and\
  \citenamefont {Babaev}}]{varney2013hierarchical}%
  \BibitemOpen
  \bibfield  {author} {\bibinfo {author} {\bibfnamefont {C.~N.}\ \bibnamefont
  {Varney}}, \bibinfo {author} {\bibfnamefont {K.~A.}\ \bibnamefont {Sellin}},
  \bibinfo {author} {\bibfnamefont {Q.-Z.}\ \bibnamefont {Wang}}, \bibinfo
  {author} {\bibfnamefont {H.}~\bibnamefont {Fangohr}}, \ and\ \bibinfo
  {author} {\bibfnamefont {E.}~\bibnamefont {Babaev}},\ }\href@noop {}
  {\bibfield  {journal} {\bibinfo  {journal} {Journal of Physics: Condensed
  Matter}\ }\textbf {\bibinfo {volume} {25}},\ \bibinfo {pages} {415702}
  (\bibinfo {year} {2013})}\BibitemShut {NoStop}%
\bibitem [{\citenamefont {Wang}(2010)}]{PhysRevB.82.132505}%
  \BibitemOpen
  \bibfield  {author} {\bibinfo {author} {\bibfnamefont {J.-P.}\ \bibnamefont
  {Wang}},\ }\href {\doibase 10.1103/PhysRevB.82.132505} {\bibfield  {journal}
  {\bibinfo  {journal} {Phys. Rev. B}\ }\textbf {\bibinfo {volume} {82}},\
  \bibinfo {pages} {132505} (\bibinfo {year} {2010})}\BibitemShut {NoStop}%
\bibitem [{\citenamefont {Drocco}\ \emph {et~al.}(2013)\citenamefont {Drocco},
  \citenamefont {Reichhardt}, \citenamefont {Reichhardt},\ and\ \citenamefont
  {Bishop}}]{Reichhardt}%
  \BibitemOpen
  \bibfield  {author} {\bibinfo {author} {\bibfnamefont {J.~A.}\ \bibnamefont
  {Drocco}}, \bibinfo {author} {\bibfnamefont {C.~J.~O.}\ \bibnamefont
  {Reichhardt}}, \bibinfo {author} {\bibfnamefont {C.}~\bibnamefont
  {Reichhardt}}, \ and\ \bibinfo {author} {\bibfnamefont {A.~R.}\ \bibnamefont
  {Bishop}},\ }\href {http://stacks.iop.org/0953-8984/25/i=34/a=345703}
  {\bibfield  {journal} {\bibinfo  {journal} {Journal of Physics: Condensed
  Matter}\ }\textbf {\bibinfo {volume} {25}},\ \bibinfo {pages} {345703}
  (\bibinfo {year} {2013})}\BibitemShut {NoStop}%
\bibitem [{\citenamefont {Meng}\ \emph {et~al.}(2014)\citenamefont {Meng},
  \citenamefont {Varney}, \citenamefont {Fangohr},\ and\ \citenamefont
  {Babaev}}]{meng2014honeycomb}%
  \BibitemOpen
  \bibfield  {author} {\bibinfo {author} {\bibfnamefont {Q.}~\bibnamefont
  {Meng}}, \bibinfo {author} {\bibfnamefont {C.~N.}\ \bibnamefont {Varney}},
  \bibinfo {author} {\bibfnamefont {H.}~\bibnamefont {Fangohr}}, \ and\
  \bibinfo {author} {\bibfnamefont {E.}~\bibnamefont {Babaev}},\ }\href@noop {}
  {\bibfield  {journal} {\bibinfo  {journal} {Physical Review B}\ }\textbf
  {\bibinfo {volume} {90}},\ \bibinfo {pages} {020509} (\bibinfo {year}
  {2014})}\BibitemShut {NoStop}%
\bibitem [{\citenamefont {Garaud}\ and\ \citenamefont
  {Babaev}(2015{\natexlab{b}})}]{Garaud.Babaev:15}%
  \BibitemOpen
  \bibfield  {author} {\bibinfo {author} {\bibfnamefont {J.}~\bibnamefont
  {Garaud}}\ and\ \bibinfo {author} {\bibfnamefont {E.}~\bibnamefont
  {Babaev}},\ }\href {\doibase 10.1103/PhysRevB.91.014510} {\bibfield
  {journal} {\bibinfo  {journal} {Phys. Rev. B}\ }\textbf {\bibinfo {volume}
  {91}},\ \bibinfo {pages} {014510} (\bibinfo {year}
  {2015}{\natexlab{b}})}\BibitemShut {NoStop}%
\bibitem [{\citenamefont {Edstr{\"o}m}(2013)}]{edstrom2013three}%
  \BibitemOpen
  \bibfield  {author} {\bibinfo {author} {\bibfnamefont {A.}~\bibnamefont
  {Edstr{\"o}m}},\ }\href@noop {} {\bibfield  {journal} {\bibinfo  {journal}
  {Physica C: Superconductivity}\ }\textbf {\bibinfo {volume} {487}},\ \bibinfo
  {pages} {19} (\bibinfo {year} {2013})}\BibitemShut {NoStop}%
\bibitem [{\citenamefont {Forgacs}\ and\ \citenamefont
  {Luk{\'a}cs}(2016)}]{forgacs2016vortices}%
  \BibitemOpen
  \bibfield  {author} {\bibinfo {author} {\bibfnamefont {P.}~\bibnamefont
  {Forgacs}}\ and\ \bibinfo {author} {\bibfnamefont {{\'A}.}~\bibnamefont
  {Luk{\'a}cs}},\ }\href@noop {} {\bibfield  {journal} {\bibinfo  {journal}
  {arXiv preprint arXiv:1603.03291}\ } (\bibinfo {year} {2016})}\BibitemShut
  {NoStop}%
\bibitem [{\citenamefont {{Gurevich}}(2003)}]{gurevich}%
  \BibitemOpen
  \bibfield  {author} {\bibinfo {author} {\bibfnamefont {A.}~\bibnamefont
  {{Gurevich}}},\ }\href {\doibase 10.1103/PhysRevB.67.184515} {\bibfield
  {journal} {\bibinfo  {journal} {\prb}\ }\textbf {\bibinfo {volume} {67}},\
  \bibinfo {eid} {184515} (\bibinfo {year} {2003})},\ \Eprint
  {http://arxiv.org/abs/cond-mat/0212129} {cond-mat/0212129} \BibitemShut
  {NoStop}%
\bibitem [{\citenamefont {Gurevich}(2007)}]{gurevich2}%
  \BibitemOpen
  \bibfield  {author} {\bibinfo {author} {\bibfnamefont {A.}~\bibnamefont
  {Gurevich}},\ }\href@noop {} {\bibfield  {journal} {\bibinfo  {journal}
  {Physica C: Superconductivity}\ }\textbf {\bibinfo {volume} {456}},\ \bibinfo
  {pages} {160} (\bibinfo {year} {2007})}\BibitemShut {NoStop}%
\bibitem [{\citenamefont {Zhitomirsky}\ and\ \citenamefont
  {Dao}(2004)}]{zhitomirsky}%
  \BibitemOpen
  \bibfield  {author} {\bibinfo {author} {\bibfnamefont {M.~E.}\ \bibnamefont
  {Zhitomirsky}}\ and\ \bibinfo {author} {\bibfnamefont {V.-H.}\ \bibnamefont
  {Dao}},\ }\href {\doibase 10.1103/PhysRevB.69.054508} {\bibfield  {journal}
  {\bibinfo  {journal} {Phys. Rev. B}\ }\textbf {\bibinfo {volume} {69}},\
  \bibinfo {pages} {054508} (\bibinfo {year} {2004})}\BibitemShut {NoStop}%
\bibitem [{\citenamefont {{Garaud}}\ \emph
  {et~al.}(2016{\natexlab{a}})\citenamefont {{Garaud}}, \citenamefont
  {{Silaev}},\ and\ \citenamefont {{Babaev}}}]{garaud2016microscopically}%
  \BibitemOpen
  \bibfield  {author} {\bibinfo {author} {\bibfnamefont {J.}~\bibnamefont
  {{Garaud}}}, \bibinfo {author} {\bibfnamefont {M.}~\bibnamefont {{Silaev}}},
  \ and\ \bibinfo {author} {\bibfnamefont {E.}~\bibnamefont {{Babaev}}},\
  }\href@noop {} {\bibfield  {journal} {\bibinfo  {journal} {ArXiv e-prints}\ }
  (\bibinfo {year} {2016}{\natexlab{a}})},\ \Eprint
  {http://arxiv.org/abs/1601.02227} {arXiv:1601.02227 [cond-mat.supr-con]}
  \BibitemShut {NoStop}%
\bibitem [{\citenamefont {Babaev}(2002)}]{frac}%
  \BibitemOpen
  \bibfield  {author} {\bibinfo {author} {\bibfnamefont {E.}~\bibnamefont
  {Babaev}},\ }\href {\doibase 10.1103/PhysRevLett.89.067001} {\bibfield
  {journal} {\bibinfo  {journal} {Phys. Rev. Lett.}\ }\textbf {\bibinfo
  {volume} {89}},\ \bibinfo {pages} {067001} (\bibinfo {year}
  {2002})}\BibitemShut {NoStop}%
\bibitem [{\citenamefont {{Speight}}(1997)}]{spe}%
  \BibitemOpen
  \bibfield  {author} {\bibinfo {author} {\bibfnamefont {J.~M.}\ \bibnamefont
  {{Speight}}},\ }\href {\doibase 10.1103/PhysRevD.55.3830} {\bibfield
  {journal} {\bibinfo  {journal} {\prd}\ }\textbf {\bibinfo {volume} {55}},\
  \bibinfo {pages} {3830} (\bibinfo {year} {1997})},\ \Eprint
  {http://arxiv.org/abs/hep-th/9603155} {hep-th/9603155} \BibitemShut {NoStop}%
\bibitem [{\citenamefont {Manton}\ and\ \citenamefont
  {Sutcliffe}(2004)}]{Manton.Sutcliffe}%
  \BibitemOpen
  \bibfield  {author} {\bibinfo {author} {\bibfnamefont {N.~S.}\ \bibnamefont
  {Manton}}\ and\ \bibinfo {author} {\bibfnamefont {P.}~\bibnamefont
  {Sutcliffe}},\ }\href@noop {} {\emph {\bibinfo {title} {{Topological
  solitons}}}}\ (\bibinfo  {publisher} {Cambridge University Press},\ \bibinfo
  {year} {2004})\ \bibinfo {note} {cambridge, UK: Univ. Pr. (2004) 493
  p}\BibitemShut {NoStop}%
\bibitem [{\citenamefont {Pearl}(1964)}]{pearl}%
  \BibitemOpen
  \bibfield  {author} {\bibinfo {author} {\bibfnamefont {J.}~\bibnamefont
  {Pearl}},\ }\href {\doibase 10.1063/1.1754056} {\bibfield  {journal}
  {\bibinfo  {journal} {Appl. Phys. Lett.}\ }\textbf {\bibinfo {volume} {5}},\
  \bibinfo {pages} {65} (\bibinfo {year} {1964})}\BibitemShut {NoStop}%
\bibitem [{\citenamefont {Bogomol'nyi}\ and\ \citenamefont
  {Vainshtein}(1976)}]{bogomol1976stability}%
  \BibitemOpen
  \bibfield  {author} {\bibinfo {author} {\bibfnamefont {E.}~\bibnamefont
  {Bogomol'nyi}}\ and\ \bibinfo {author} {\bibfnamefont {A.}~\bibnamefont
  {Vainshtein}},\ }\href@noop {} {\bibfield  {journal} {\bibinfo  {journal}
  {Sov. J. Nucl. Phys.(Engl. Transl.);(United States)}\ }\textbf {\bibinfo
  {volume} {23}} (\bibinfo {year} {1976})}\BibitemShut {NoStop}%
\bibitem [{\citenamefont {Saint-James}\ \emph {et~al.}(1969)\citenamefont
  {Saint-James}, \citenamefont {Sarma},\ and\ \citenamefont
  {Thomas}}]{saint1969type}%
  \BibitemOpen
  \bibfield  {author} {\bibinfo {author} {\bibfnamefont {D.}~\bibnamefont
  {Saint-James}}, \bibinfo {author} {\bibfnamefont {G.}~\bibnamefont {Sarma}},
  \ and\ \bibinfo {author} {\bibfnamefont {E.~J.}\ \bibnamefont {Thomas}},\
  }\href@noop {} {\emph {\bibinfo {title} {TYPE-II SUPERCONDUCTIVITY.}}},\
  \bibinfo {type} {Tech. Rep.}\ (\bibinfo  {institution} {CEN, Saclay,
  France},\ \bibinfo {year} {1969})\BibitemShut {NoStop}%
\bibitem [{\citenamefont {Shifman}(2012)}]{shifman2012advanced}%
  \BibitemOpen
  \bibfield  {author} {\bibinfo {author} {\bibfnamefont {M.}~\bibnamefont
  {Shifman}},\ }\href {https://books.google.se/books?id=zeQuWycXV3oC} {\emph
  {\bibinfo {title} {Advanced Topics in Quantum Field Theory: A Lecture
  Course}}}\ (\bibinfo  {publisher} {Cambridge University Press},\ \bibinfo
  {year} {2012})\BibitemShut {NoStop}%
\bibitem [{\citenamefont {Svistunov}\ \emph {et~al.}(2015)\citenamefont
  {Svistunov}, \citenamefont {Babaev},\ and\ \citenamefont {Prokof'ev}}]{ssm}%
  \BibitemOpen
  \bibfield  {author} {\bibinfo {author} {\bibfnamefont {B.~V.}\ \bibnamefont
  {Svistunov}}, \bibinfo {author} {\bibfnamefont {E.~S.}\ \bibnamefont
  {Babaev}}, \ and\ \bibinfo {author} {\bibfnamefont {N.~V.}\ \bibnamefont
  {Prokof'ev}},\ }\href@noop {} {\emph {\bibinfo {title} {Superfluid states of
  matter}}}\ (\bibinfo  {publisher} {Crc Press},\ \bibinfo {year}
  {2015})\BibitemShut {NoStop}%
\bibitem [{\citenamefont {Jacobs}(1973)}]{jacobs}%
  \BibitemOpen
  \bibfield  {author} {\bibinfo {author} {\bibfnamefont {A.}~\bibnamefont
  {Jacobs}},\ }\href {\doibase 10.1007/BF00655246} {\bibfield  {journal}
  {\bibinfo  {journal} {J. Low Temp. Phys.}\ }\textbf {\bibinfo {volume}
  {10}},\ \bibinfo {pages} {137} (\bibinfo {year} {1973})}\BibitemShut
  {NoStop}%
\bibitem [{\citenamefont {{Leung}}\ and\ \citenamefont
  {{Jacobs}}(1973)}]{leung1}%
  \BibitemOpen
  \bibfield  {author} {\bibinfo {author} {\bibfnamefont {M.~C.}\ \bibnamefont
  {{Leung}}}\ and\ \bibinfo {author} {\bibfnamefont {A.~E.}\ \bibnamefont
  {{Jacobs}}},\ }\href {\doibase 10.1007/BF00656560} {\bibfield  {journal}
  {\bibinfo  {journal} {J. Low Temp. Phys.}\ }\textbf {\bibinfo {volume}
  {11}},\ \bibinfo {pages} {395} (\bibinfo {year} {1973})}\BibitemShut
  {NoStop}%
\bibitem [{\citenamefont {{Eilenberger}}\ and\ \citenamefont
  {{B{\"u}ttner}}(1969)}]{eilenberger}%
  \BibitemOpen
  \bibfield  {author} {\bibinfo {author} {\bibfnamefont {G.}~\bibnamefont
  {{Eilenberger}}}\ and\ \bibinfo {author} {\bibfnamefont {H.}~\bibnamefont
  {{B{\"u}ttner}}},\ }\href {\doibase 10.1007/BF01393061} {\bibfield  {journal}
  {\bibinfo  {journal} {Zeitschrift fur Physik}\ }\textbf {\bibinfo {volume}
  {224}},\ \bibinfo {pages} {335} (\bibinfo {year} {1969})}\BibitemShut
  {NoStop}%
\bibitem [{\citenamefont {Jacobs}(1971)}]{PhysRevB.4.3029}%
  \BibitemOpen
  \bibfield  {author} {\bibinfo {author} {\bibfnamefont {A.~E.}\ \bibnamefont
  {Jacobs}},\ }\href {\doibase 10.1103/PhysRevB.4.3029} {\bibfield  {journal}
  {\bibinfo  {journal} {Phys. Rev. B}\ }\textbf {\bibinfo {volume} {4}},\
  \bibinfo {pages} {3029} (\bibinfo {year} {1971})}\BibitemShut {NoStop}%
\bibitem [{\citenamefont {Ovchinnikov}(1999)}]{ovchinnikov1999generalized}%
  \BibitemOpen
  \bibfield  {author} {\bibinfo {author} {\bibfnamefont {Y.~N.}\ \bibnamefont
  {Ovchinnikov}},\ }\href@noop {} {\bibfield  {journal} {\bibinfo  {journal}
  {Journal of Experimental and Theoretical Physics}\ }\textbf {\bibinfo
  {volume} {88}},\ \bibinfo {pages} {398} (\bibinfo {year} {1999})}\BibitemShut
  {NoStop}%
\bibitem [{\citenamefont {Ovchinnikov}(2013)}]{Ovchinnikov2013}%
  \BibitemOpen
  \bibfield  {author} {\bibinfo {author} {\bibfnamefont {Y.~N.}\ \bibnamefont
  {Ovchinnikov}},\ }\href {\doibase 10.1134/S1063776113110046} {\bibfield
  {journal} {\bibinfo  {journal} {Journal of Experimental and Theoretical
  Physics}\ }\textbf {\bibinfo {volume} {117}},\ \bibinfo {pages} {480}
  (\bibinfo {year} {2013})}\BibitemShut {NoStop}%
\bibitem [{\citenamefont {{Meng}}\ \emph {et~al.}(2016)\citenamefont {{Meng}},
  \citenamefont {{Varney}}, \citenamefont {{Fangohr}},\ and\ \citenamefont
  {{Babaev}}}]{2016arXiv160500524M}%
  \BibitemOpen
  \bibfield  {author} {\bibinfo {author} {\bibfnamefont {Q.}~\bibnamefont
  {{Meng}}}, \bibinfo {author} {\bibfnamefont {C.~N.}\ \bibnamefont
  {{Varney}}}, \bibinfo {author} {\bibfnamefont {H.}~\bibnamefont {{Fangohr}}},
  \ and\ \bibinfo {author} {\bibfnamefont {E.}~\bibnamefont {{Babaev}}},\
  }\href@noop {} {\bibfield  {journal} {\bibinfo  {journal} {ArXiv e-prints}\ }
  (\bibinfo {year} {2016})},\ \Eprint {http://arxiv.org/abs/1605.00524}
  {arXiv:1605.00524 [cond-mat.supr-con]} \BibitemShut {NoStop}%
\bibitem [{\citenamefont {{Diaz-Mendez}}\ \emph {et~al.}(2016)\citenamefont
  {{Diaz-Mendez}}, \citenamefont {{Mezzacapo}}, \citenamefont {{Lechner}},
  \citenamefont {{Cinti}}, \citenamefont {{Babaev}},\ and\ \citenamefont
  {{Pupillo}}}]{2016arXiv160500553D}%
  \BibitemOpen
  \bibfield  {author} {\bibinfo {author} {\bibfnamefont {R.}~\bibnamefont
  {{Diaz-Mendez}}}, \bibinfo {author} {\bibfnamefont {F.}~\bibnamefont
  {{Mezzacapo}}}, \bibinfo {author} {\bibfnamefont {W.}~\bibnamefont
  {{Lechner}}}, \bibinfo {author} {\bibfnamefont {F.}~\bibnamefont {{Cinti}}},
  \bibinfo {author} {\bibfnamefont {E.}~\bibnamefont {{Babaev}}}, \ and\
  \bibinfo {author} {\bibfnamefont {G.}~\bibnamefont {{Pupillo}}},\ }\href@noop
  {}  {\bibfield  {journal} {\bibinfo  {journal} {Physical Review
 Letters}\ }\textbf {\bibinfo {volume} {118}},\ \bibinfo {pages} {067001}
  (\bibinfo {year} {2017})} \BibitemShut {NoStop}%
\bibitem [{\citenamefont {Sm{\o}rgrav}\ \emph {et~al.}(2005)\citenamefont
  {Sm{\o}rgrav}, \citenamefont {Smiseth}, \citenamefont {Babaev},\ and\
  \citenamefont {Sudb{\o}}}]{sublattice}%
  \BibitemOpen
  \bibfield  {author} {\bibinfo {author} {\bibfnamefont {E.}~\bibnamefont
  {Sm{\o}rgrav}}, \bibinfo {author} {\bibfnamefont {J.}~\bibnamefont
  {Smiseth}}, \bibinfo {author} {\bibfnamefont {E.}~\bibnamefont {Babaev}}, \
  and\ \bibinfo {author} {\bibfnamefont {A.}~\bibnamefont {Sudb{\o}}},\
  }\href@noop {} {\bibfield  {journal} {\bibinfo  {journal} {Physical Review
 Letters}\ }\textbf {\bibinfo {volume} {94}},\ \bibinfo {pages} {96401}
  (\bibinfo {year} {2005})}\BibitemShut {NoStop}%
\bibitem [{\citenamefont {Ng}\ and\ \citenamefont {Nagaosa}(2009)}]{Ng}%
  \BibitemOpen
  \bibfield  {author} {\bibinfo {author} {\bibfnamefont {T.~K.}\ \bibnamefont
  {Ng}}\ and\ \bibinfo {author} {\bibfnamefont {N.}~\bibnamefont {Nagaosa}},\
  }\href {\doibase 10.1209/0295-5075/87/17003} {\bibfield  {journal} {\bibinfo
  {journal} {Europhysics Letters}\ }\textbf {\bibinfo {volume} {87}},\ \bibinfo
  {pages} {17003} (\bibinfo {year} {2009})}\BibitemShut {NoStop}%
\bibitem [{\citenamefont {Lin}\ and\ \citenamefont {Hu}(2012)}]{Lin}%
  \BibitemOpen
  \bibfield  {author} {\bibinfo {author} {\bibfnamefont {S.-Z.}\ \bibnamefont
  {Lin}}\ and\ \bibinfo {author} {\bibfnamefont {X.}~\bibnamefont {Hu}},\
  }\href {\doibase 10.1103/PhysRevLett.108.177005} {\bibfield  {journal}
  {\bibinfo  {journal} {Phys. Rev. Lett.}\ }\textbf {\bibinfo {volume} {108}},\
  \bibinfo {pages} {177005} (\bibinfo {year} {2012})}\BibitemShut {NoStop}%
\bibitem [{\citenamefont {Garaud}\ \emph {et~al.}(2011)\citenamefont {Garaud},
  \citenamefont {Carlstr\"om},\ and\ \citenamefont
  {Babaev}}]{Garaud.Carlstrom.ea:11}%
  \BibitemOpen
  \bibfield  {author} {\bibinfo {author} {\bibfnamefont {J.}~\bibnamefont
  {Garaud}}, \bibinfo {author} {\bibfnamefont {J.}~\bibnamefont {Carlstr\"om}},
  \ and\ \bibinfo {author} {\bibfnamefont {E.}~\bibnamefont {Babaev}},\ }\href
  {\doibase 10.1103/PhysRevLett.107.197001} {\bibfield  {journal} {\bibinfo
  {journal} {Phys. Rev. Lett.}\ }\textbf {\bibinfo {volume} {107}},\ \bibinfo
  {pages} {197001} (\bibinfo {year} {2011})}\BibitemShut {NoStop}%
\bibitem [{\citenamefont {Garaud}\ \emph {et~al.}(2013)\citenamefont {Garaud},
  \citenamefont {Carlstr\"om}, \citenamefont {Babaev},\ and\ \citenamefont
  {Speight}}]{Garaud.Carlstrom.ea:13}%
  \BibitemOpen
  \bibfield  {author} {\bibinfo {author} {\bibfnamefont {J.}~\bibnamefont
  {Garaud}}, \bibinfo {author} {\bibfnamefont {J.}~\bibnamefont {Carlstr\"om}},
  \bibinfo {author} {\bibfnamefont {E.}~\bibnamefont {Babaev}}, \ and\ \bibinfo
  {author} {\bibfnamefont {M.}~\bibnamefont {Speight}},\ }\href {\doibase
  10.1103/PhysRevB.87.014507} {\bibfield  {journal} {\bibinfo  {journal} {Phys.
  Rev. B}\ }\textbf {\bibinfo {volume} {87}},\ \bibinfo {pages} {014507}
  (\bibinfo {year} {2013})}\BibitemShut {NoStop}%
\bibitem [{\citenamefont {Garaud}\ and\ \citenamefont
  {Babaev}(2014)}]{Garaud.Babaev:14}%
  \BibitemOpen
  \bibfield  {author} {\bibinfo {author} {\bibfnamefont {J.}~\bibnamefont
  {Garaud}}\ and\ \bibinfo {author} {\bibfnamefont {E.}~\bibnamefont
  {Babaev}},\ }\href {\doibase 10.1103/PhysRevLett.112.017003} {\bibfield
  {journal} {\bibinfo  {journal} {Phys. Rev. Lett.}\ }\textbf {\bibinfo
  {volume} {112}},\ \bibinfo {pages} {017003} (\bibinfo {year}
  {2014})}\BibitemShut {NoStop}%
\bibitem [{\citenamefont {Weston}\ and\ \citenamefont
  {Babaev}(2013)}]{Weston.Babaev:13}%
  \BibitemOpen
  \bibfield  {author} {\bibinfo {author} {\bibfnamefont {D.}~\bibnamefont
  {Weston}}\ and\ \bibinfo {author} {\bibfnamefont {E.}~\bibnamefont
  {Babaev}},\ }\href {\doibase 10.1103/PhysRevB.88.214507} {\bibfield
  {journal} {\bibinfo  {journal} {Phys. Rev. B}\ }\textbf {\bibinfo {volume}
  {88}},\ \bibinfo {pages} {214507} (\bibinfo {year} {2013})}\BibitemShut
  {NoStop}%
\bibitem [{\citenamefont {Carlstr\"om}\ and\ \citenamefont
  {Babaev}(2015)}]{carlstrom2014spontaneous}%
  \BibitemOpen
  \bibfield  {author} {\bibinfo {author} {\bibfnamefont {J.}~\bibnamefont
  {Carlstr\"om}}\ and\ \bibinfo {author} {\bibfnamefont {E.}~\bibnamefont
  {Babaev}},\ }\href {\doibase 10.1103/PhysRevB.91.140504} {\bibfield
  {journal} {\bibinfo  {journal} {Phys. Rev. B}\ }\textbf {\bibinfo {volume}
  {91}},\ \bibinfo {pages} {140504} (\bibinfo {year} {2015})}\BibitemShut
  {NoStop}%
\bibitem [{\citenamefont {Silaev}\ and\ \citenamefont
  {Babaev}(2013)}]{Silaevvisc}%
  \BibitemOpen
  \bibfield  {author} {\bibinfo {author} {\bibfnamefont {M.}~\bibnamefont
  {Silaev}}\ and\ \bibinfo {author} {\bibfnamefont {E.}~\bibnamefont
  {Babaev}},\ }\href {\doibase 10.1103/PhysRevB.88.220504} {\bibfield
  {journal} {\bibinfo  {journal} {Phys. Rev. B}\ }\textbf {\bibinfo {volume}
  {88}},\ \bibinfo {pages} {220504} (\bibinfo {year} {2013})}\BibitemShut
  {NoStop}%
\bibitem [{\citenamefont {Silaev}\ \emph {et~al.}(2015)\citenamefont {Silaev},
  \citenamefont {Garaud},\ and\ \citenamefont {Babaev}}]{Silaev.Garaud.ea:15}%
  \BibitemOpen
  \bibfield  {author} {\bibinfo {author} {\bibfnamefont {M.}~\bibnamefont
  {Silaev}}, \bibinfo {author} {\bibfnamefont {J.}~\bibnamefont {Garaud}}, \
  and\ \bibinfo {author} {\bibfnamefont {E.}~\bibnamefont {Babaev}},\ }\href
  {\doibase 10.1103/PhysRevB.92.174510} {\bibfield  {journal} {\bibinfo
  {journal} {Phys. Rev. B}\ }\textbf {\bibinfo {volume} {92}},\ \bibinfo
  {pages} {174510} (\bibinfo {year} {2015})}\BibitemShut {NoStop}%
\bibitem [{\citenamefont {{Garaud}}\ \emph
  {et~al.}(2016{\natexlab{b}})\citenamefont {{Garaud}}, \citenamefont
  {{Silaev}},\ and\ \citenamefont {{Babaev}}}]{2016PhRvL.116i7002G}%
  \BibitemOpen
  \bibfield  {author} {\bibinfo {author} {\bibfnamefont {J.}~\bibnamefont
  {{Garaud}}}, \bibinfo {author} {\bibfnamefont {M.}~\bibnamefont {{Silaev}}},
  \ and\ \bibinfo {author} {\bibfnamefont {E.}~\bibnamefont {{Babaev}}},\
  }\href {\doibase 10.1103/PhysRevLett.116.097002} {\bibfield  {journal}
  {\bibinfo  {journal} {Physical Review Letters}\ }\textbf {\bibinfo {volume}
  {116}},\ \bibinfo {eid} {097002} (\bibinfo {year} {2016}{\natexlab{b}})},\
  \Eprint {http://arxiv.org/abs/1507.04712} {arXiv:1507.04712
  [cond-mat.supr-con]} \BibitemShut {NoStop}%
\bibitem [{\citenamefont {Halperin}\ \emph {et~al.}(1974)\citenamefont
  {Halperin}, \citenamefont {Lubensky},\ and\ \citenamefont
  {Ma}}]{HalperinLubenskyMa}%
  \BibitemOpen
  \bibfield  {author} {\bibinfo {author} {\bibfnamefont {B.~I.}\ \bibnamefont
  {Halperin}}, \bibinfo {author} {\bibfnamefont {T.~C.}\ \bibnamefont
  {Lubensky}}, \ and\ \bibinfo {author} {\bibfnamefont {S.-K.}\ \bibnamefont
  {Ma}},\ }\href {\doibase 10.1103/PhysRevLett.32.292} {\bibfield  {journal}
  {\bibinfo  {journal} {Phys. Rev. Lett.}\ }\textbf {\bibinfo {volume} {32}},\
  \bibinfo {pages} {292} (\bibinfo {year} {1974})}\BibitemShut {NoStop}%
\bibitem [{\citenamefont {Coleman}\ and\ \citenamefont
  {Weinberg}(1973)}]{COLEMANWEINBERG}%
  \BibitemOpen
  \bibfield  {author} {\bibinfo {author} {\bibfnamefont {S.}~\bibnamefont
  {Coleman}}\ and\ \bibinfo {author} {\bibfnamefont {E.}~\bibnamefont
  {Weinberg}},\ }\href {\doibase 10.1103/PhysRevD.7.1888} {\bibfield  {journal}
  {\bibinfo  {journal} {Phys. Rev. D}\ }\textbf {\bibinfo {volume} {7}},\
  \bibinfo {pages} {1888} (\bibinfo {year} {1973})}\BibitemShut {NoStop}%
\bibitem [{\citenamefont {Dasgupta}\ and\ \citenamefont
  {Halperin}(1981)}]{DasguptaHalperin}%
  \BibitemOpen
  \bibfield  {author} {\bibinfo {author} {\bibfnamefont {C.}~\bibnamefont
  {Dasgupta}}\ and\ \bibinfo {author} {\bibfnamefont {B.~I.}\ \bibnamefont
  {Halperin}},\ }\href {\doibase 10.1103/PhysRevLett.47.1556} {\bibfield
  {journal} {\bibinfo  {journal} {Phys. Rev. Lett.}\ }\textbf {\bibinfo
  {volume} {47}},\ \bibinfo {pages} {1556} (\bibinfo {year}
  {1981})}\BibitemShut {NoStop}%
\bibitem [{\citenamefont {Peskin}(1978)}]{Peskin1978122}%
  \BibitemOpen
  \bibfield  {author} {\bibinfo {author} {\bibfnamefont {M.~E.}\ \bibnamefont
  {Peskin}},\ }\href {\doibase http://dx.doi.org/10.1016/0003-4916(78)90252-X}
  {\bibfield  {journal} {\bibinfo  {journal} {Annals of Physics}\ }\textbf
  {\bibinfo {volume} {113}},\ \bibinfo {pages} {122 } (\bibinfo {year}
  {1978})}\BibitemShut {NoStop}%
\bibitem [{\citenamefont {Thomas}\ and\ \citenamefont
  {Stone}(1978)}]{Thomas1978513}%
  \BibitemOpen
  \bibfield  {author} {\bibinfo {author} {\bibfnamefont {P.~R.}\ \bibnamefont
  {Thomas}}\ and\ \bibinfo {author} {\bibfnamefont {M.}~\bibnamefont {Stone}},\
  }\href {\doibase http://dx.doi.org/10.1016/0550-3213(78)90383-8} {\bibfield
  {journal} {\bibinfo  {journal} {Nuclear Physics B}\ }\textbf {\bibinfo
  {volume} {144}},\ \bibinfo {pages} {513 } (\bibinfo {year}
  {1978})}\BibitemShut {NoStop}%
\bibitem [{\citenamefont {Bartholomew}(1983)}]{bartol}%
  \BibitemOpen
  \bibfield  {author} {\bibinfo {author} {\bibfnamefont {J.}~\bibnamefont
  {Bartholomew}},\ }\href {\doibase 10.1103/PhysRevB.28.5378} {\bibfield
  {journal} {\bibinfo  {journal} {Phys. Rev. B}\ }\textbf {\bibinfo {volume}
  {28}},\ \bibinfo {pages} {5378} (\bibinfo {year} {1983})}\BibitemShut
  {NoStop}%
\bibitem [{\citenamefont {Mo}\ \emph {et~al.}(2002)\citenamefont {Mo},
  \citenamefont {Hove},\ and\ \citenamefont {Sudb\o{}}}]{SudboKappa}%
  \BibitemOpen
  \bibfield  {author} {\bibinfo {author} {\bibfnamefont {S.}~\bibnamefont
  {Mo}}, \bibinfo {author} {\bibfnamefont {J.}~\bibnamefont {Hove}}, \ and\
  \bibinfo {author} {\bibfnamefont {A.}~\bibnamefont {Sudb\o{}}},\ }\href
  {\doibase 10.1103/PhysRevB.65.104501} {\bibfield  {journal} {\bibinfo
  {journal} {Phys. Rev. B}\ }\textbf {\bibinfo {volume} {65}},\ \bibinfo
  {pages} {104501} (\bibinfo {year} {2002})}\BibitemShut {NoStop}%
\bibitem [{\citenamefont {Hove}\ \emph {et~al.}(2002)\citenamefont {Hove},
  \citenamefont {Mo},\ and\ \citenamefont {Sudb\o{}}}]{SudboKappa2}%
  \BibitemOpen
  \bibfield  {author} {\bibinfo {author} {\bibfnamefont {J.}~\bibnamefont
  {Hove}}, \bibinfo {author} {\bibfnamefont {S.}~\bibnamefont {Mo}}, \ and\
  \bibinfo {author} {\bibfnamefont {A.}~\bibnamefont {Sudb\o{}}},\ }\href
  {\doibase 10.1103/PhysRevB.66.064524} {\bibfield  {journal} {\bibinfo
  {journal} {Phys. Rev. B}\ }\textbf {\bibinfo {volume} {66}},\ \bibinfo
  {pages} {064524} (\bibinfo {year} {2002})}\BibitemShut {NoStop}%
\bibitem [{\citenamefont {Meier}\ \emph {et~al.}(2015)\citenamefont {Meier},
  \citenamefont {Babaev},\ and\ \citenamefont {Wallin}}]{meier}%
  \BibitemOpen
  \bibfield  {author} {\bibinfo {author} {\bibfnamefont {H.}~\bibnamefont
  {Meier}}, \bibinfo {author} {\bibfnamefont {E.}~\bibnamefont {Babaev}}, \
  and\ \bibinfo {author} {\bibfnamefont {M.}~\bibnamefont {Wallin}},\
  }\href@noop {} {\bibfield  {journal} {\bibinfo  {journal} {Physical Review
  B}\ }\textbf {\bibinfo {volume} {91}},\ \bibinfo {pages} {094508} (\bibinfo
  {year} {2015})}\BibitemShut {NoStop}%
\end{thebibliography}
%

\end{document}